\documentclass[%
 reprint,
 amsmath,amssymb,superscriptaddress,
 aps,
 prd,
 floatfix
]{revtex4-1}

\usepackage{graphicx}
\usepackage{float}
\usepackage{physics}
\usepackage[dvipsnames]{xcolor}
\usepackage[shortlabels]{enumitem}

\usepackage[normalem]{ulem}

\allowdisplaybreaks
\begin{document}

\title{High-precision baryon number cumulants from lattice QCD in a finite box: \\
cumulant ratios, Lee-Yang zeros and critical endpoint predictions}

\author{Alexander Adam}
\affiliation{Department of Physics, Wuppertal University, Gaussstr.  20, D-42119, Wuppertal, Germany}

\author{Szabolcs Bors\'anyi}
\affiliation{Department of Physics, Wuppertal University, Gaussstr.  20, D-42119, Wuppertal, Germany}

\author{Zolt\'an Fodor}
\affiliation{Pennsylvania State University, Department of Physics, State College, PA 16801, USA}
\affiliation{Department of Physics, Wuppertal University, Gaussstr.  20, D-42119, Wuppertal, Germany}
\affiliation{Institute  for Theoretical Physics, ELTE E\"otv\"os Lor\' and University, P\'azm\'any P. s\'et\'any 1/A, H-1117 Budapest, Hungary}
\affiliation{J\"ulich Supercomputing Centre, Forschungszentrum J\"ulich, D-52425 J\"ulich, Germany}

\author{Jana N. Guenther}
\affiliation{Department of Physics, Wuppertal University, Gaussstr.  20, D-42119, Wuppertal, Germany}

\author{Piyush Kumar}
\affiliation{Department of Physics, Wuppertal University, Gaussstr.  20, D-42119, Wuppertal, Germany}

\author{Paolo Parotto}
\affiliation{Dipartimento di Fisica, Universit\`a di Torino and INFN Torino, Via P. Giuria 1, I-10125 Torino, Italy}

\author{Attila P\'asztor}
\affiliation{Institute  for Theoretical Physics, ELTE E\"otv\"os Lor\' and University, P\'azm\'any P. s\'et\'any 1/A, H-1117 Budapest, Hungary}

\author{Chik Him Wong}
\affiliation{Department of Physics, Wuppertal University, Gaussstr.  20, D-42119, Wuppertal, Germany}

\date{\today}

\begin{abstract}
We have performed high-statistics lattice simulations using 4HEX improved staggered fermions on $16^3 \times 8$ lattices. We calculated
the Taylor expansion coefficients of the pressure with respect to the baryochemical potential to the tenth order 
at zero, and fourth order at purely imaginary chemical potentials.
We used this data to construct rational function approximations of the
free energy. We use a rational ansatz that explicitly
satisfies the charge conjugation symmetry and the Roberge-Weiss periodicity, which are exact
properties of the QCD free energy. 
We use this ansatz to estimate the position of Lee-Yang zeros in the complex chemical potential plane. The temperature dependence of the
imaginary part of the Lee-Yang zeros is then fitted with ansätze motivated by the universal behavior of the free energy near a 3D Ising critical point. In principle, this allows one to estimate the temperature of the critical endpoint. 
We consider several sources of systematic errors.
On this single lattice spacing we find that
with $84\%$ probability, the chiral critical endpoint is either
below $103$~MeV temperature or it does not exist.
We also identify some caveats of the method, which do not
disappear even with the extremely high statistics of this present study. 
We discuss to what extent these can be eliminated by future high statistics lattice analyses.
\end{abstract}

\maketitle

\section{Introduction}
A major goal of heavy-ion physics research is the exploration of the 
phase diagram of strongly interacting matter in the temperature($T$)-baryon density (or baryochemical potential $\mu_B$) plane.  
Although it is well established that the transition between quarks
and hadrons at $\mu_B=0$ is a smooth crossover~\cite{Aoki:2006we}, our knowledge of the
phase diagram at $\mu_B>0$ is limited. 

First-principles lattice QCD calculations are hampered by the sign problem, 
severely limiting the region in $\mu_B$ where results are
available. Some well-established quantities determined with a lattice QCD approach include
the curvature of the phase diagram at $\mu_B=0$~\cite{Bonati:2018nut, HotQCD:2018pds, Borsanyi:2020fev} and the derivatives of the pressure with
respect to the chemical potentials~\cite{Bollweg:2022rps, Borsanyi:2023wno, Borsanyi:2018grb}.

Higher chemical potentials are explored with chiral effective model calculations ~\cite{ Kovacs:2016juc, Fu:2023lcm}, truncated Dyson-Schwinger equations ~\cite{Isserstedt:2019pgx, Gao:2020qsj} and holography-based modeling~\cite{Hippert:2023bel}. These approaches tend to predict a critical
endpoint where the crossover line becomes a line first-order transition, at higher
values of the baryochemical potential.

Recently, the lattice QCD community has developed
extrapolation schemes that can reach higher values of the chemical potentials and thus shed some light on the position of the coveted
critical endpoint. These extrpaolations employ lattice QCD data
collected either at zero or purely imaginary chemical potentials. A general
property of these methods is that they are based on ans\"atze for
the free energy that are not polynomials in the chemical potential. Such a non-polynomial structure is necessary to capture the
onset of a true phase transition, which leads to
non-analytic behavior in the free energy. 

Two related approaches include the $T'$  expansion ~\cite{Borsanyi:2021sxv, Borsanyi:2022qlh, Kahangirwe:2024xyl, Wen:2024hbz, Abuali:2025tbd} and 
the extrapolation of contours of constant entropy ~\cite{Shah:2024img, Borsanyi:2025dyp, Marczenko:2025znt}. The current state-of-the-art using
this approach is a $2\sigma$ lower bound on the position of the critical endpoint
at $\mu_B\approx450$MeV.

In this manuscript we study Lee-Yang zeros, which offer a different approach to the study of the phase diagram~\cite{Fodor:2004nz, Giordano:2019slo, Giordano:2019gev, Dimopoulos:2021vrk, Bollweg:2022rps, Basar:2023nkp, Clarke:2024ugt}, by means of a
two-step extrapolation (see Ref.~\cite{Skokov:2024fac} for a detailed explanation). In the first step, one extrapolates to the complex $\mu_B$ plane at a fixed value of the temperature $T$. The goal of this first step is to estimate the position of Lee-Yang zeros. These are zeros of 
the partition function, logarithmic branch points of the free energy and poles of the 
baryon number and baryon number susceptibility. At temperatures where 
the transition at $\mu_B>0$ is a crossover, the position of these zeros has
a non-vanishing imaginary part. In the second step of the approach, one then 
attempts to extrapolate the value of this imaginary part to lower temperatures,
to estimate where it becomes zero. This provides an estimate of the
critical endpoint (CEP) position. Recent CEP estimates using this approach include Refs.~\cite{Basar:2023nkp, Clarke:2024ugt}. 

Both steps of this two-step approach can potentially suffer from large systematic
uncertainties. The main goal of this manuscript is to estimate the uncertainties
of both of these steps, and to see how they propagate to the final predictions
on the critical endpoint position, thus testing the reliability of the approach.
We will show that the systematics, especially of the second extrapolation, are
quite substantial. This means that instead of a sharp prediction on the 
CEP position, the extrapolations only give an upper bound on the temperature
of the critical endpoint. We conduct this study in a setting where we have very accurate input 
data: in a finite volume and at finite lattice spacing. This means that inevitably
our results are subject to volume and cut-off effects. We 
cannot yet provide an estimate of the systematic errors related 
to these effects, and leave that for future work.

The structure of this manuscript is as follows. We explain the lattice setup in the next section. In section 3 we present our results for the baryon chemical
potential derivatives we will use in the subsequent analysis of Lee-Yang
zeros. This section also includes simple - Taylor expansion based - estimates of the phenomenologically relevant fourth to second and third to first cumulant ratios at non-zero $\mu_B$. In section 4 we present our analysis of the Lee-Yang zeros based on a rational function
approximation and the subsequent extrapolation
of the CEP position. We will emphasize the source of different systematic errors
and finally show a posterior distribution for the estimated CEP position. Our results include a systematic error estimate of the two different extrapolation steps. However, we note that, at present, potentially one of the
most important sources of systematic uncertainties -- namely the temperature range
used for the extrapolation to the critical point -- cannot be estimated, due to the limited range
in the available lattice data. This is a feature that
our study shares with other studies available
in the literature.

\begin{figure}[t]
    \centering
    \includegraphics[width=0.8\linewidth]{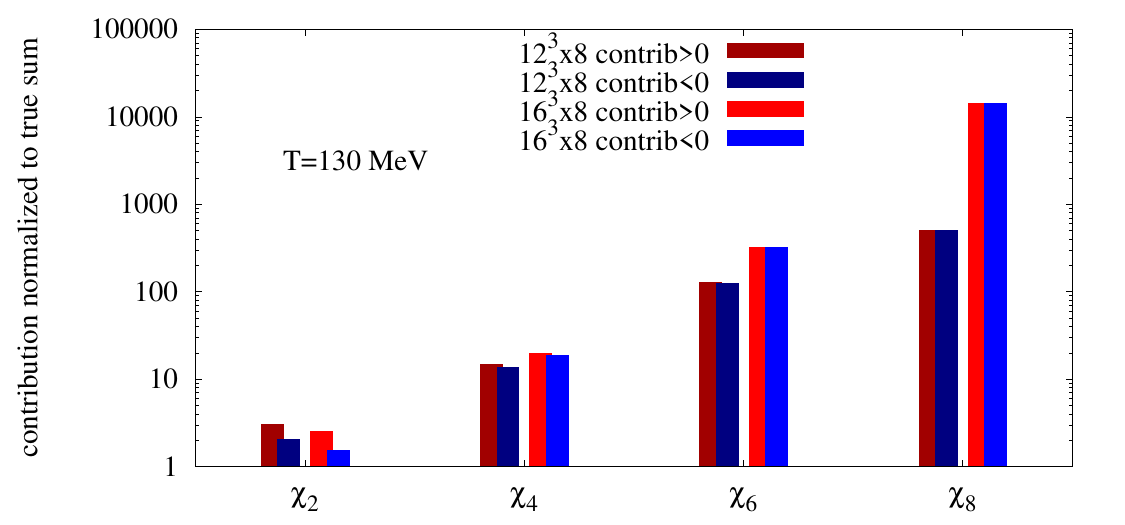}
    \caption{
    \label{fig:cancellations}
    Cancellations between positive and negative terms for the Taylor coefficients $\chi_n$ for up to the eighth order at $T=130$~MeV. Positive contributions are shown in red, while negative contributions are shown in blue. A smaller volume ($12^3 \times 8$)
    is also shown, to illustrate the volume dependence of the 
    cancellations.
}
\end{figure}

\begin{figure*}[t]
    \centering
    \includegraphics[width=0.321\linewidth]{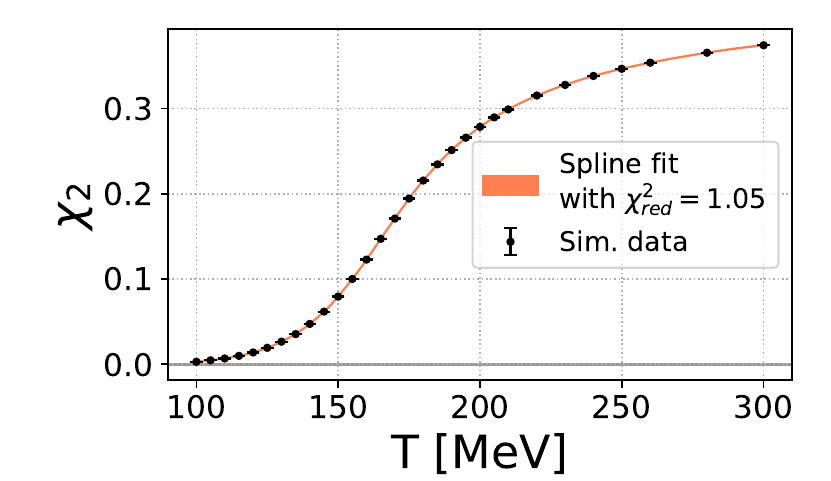}
    \includegraphics[width=0.321\linewidth]{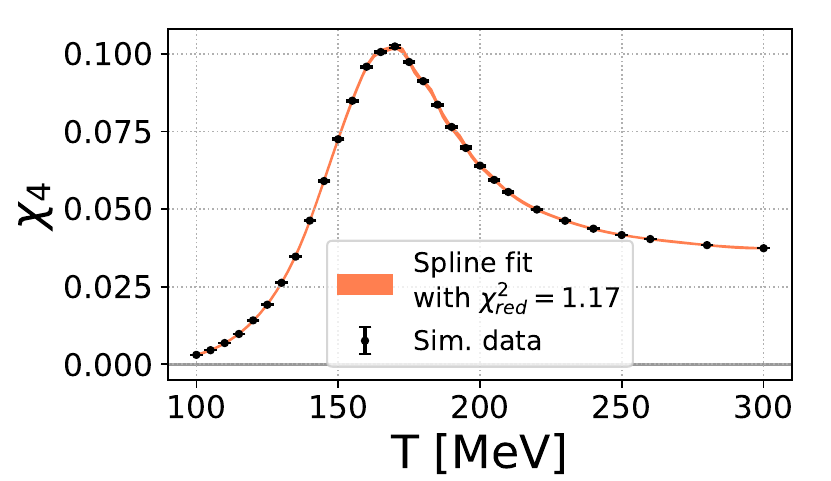}
    \includegraphics[width=0.321\linewidth]{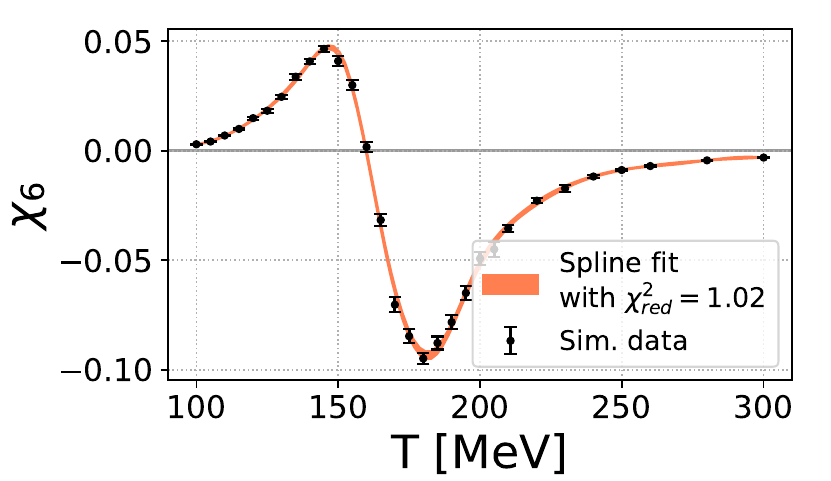}
    \includegraphics[width=0.420\linewidth]{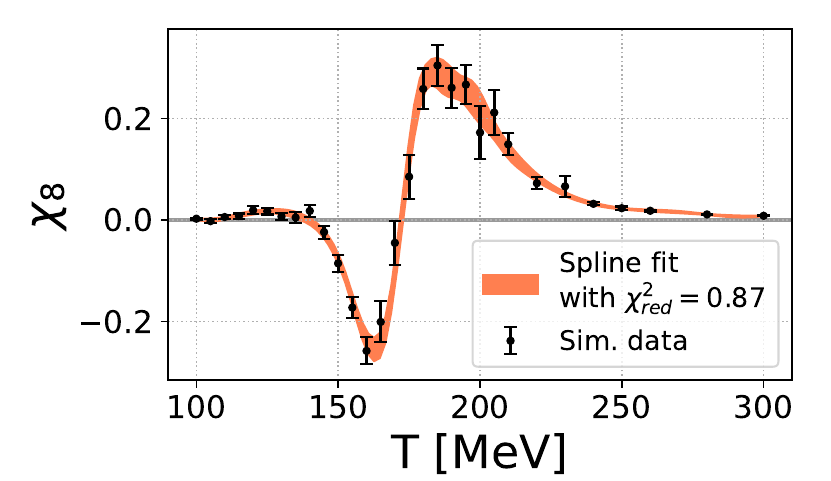}
    \includegraphics[width=0.420\linewidth]{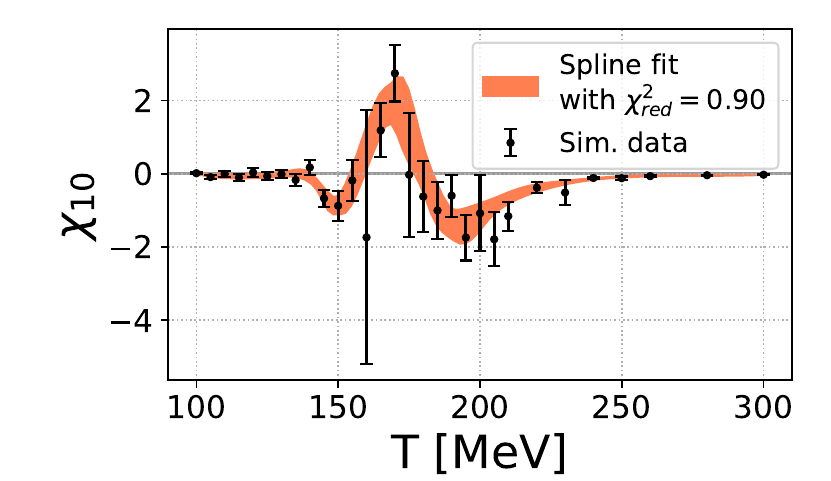}
    \caption{
    \label{fig:chis}
    Derivatives of the pressure with respect to the baryo-chemical potential
    at $\mu_B=0$ as functions of the temperature $T$ on our $16^3 \times 8$ 4HEX ensembles, together with a spline interpolation of the data points.
}
\end{figure*}

\section{Baryon number fluctuations from Lattice QCD}

\subsection{Lattice observables and simulation details}

We use a fixed lattice size $16^3 \times 8$, meaning that our
physical volume is fixed in temperature units to $LT=2$.
We use ensembles with 4HEX staggered fermions as in
our recent works  \cite{Borsanyi:2023wno} and 
\cite{Borsanyi:2025lim}, but the statistics was 
increased considerably to around 2 million configurations per
temperature. Furthermore, we also simulated lattices at eight imaginary
values of the chemical potential, with a statistics of around 100,000
configurations per temperature and chemical potential (see statistics in Table \ref{tab:stat} in Appendix \ref{app:stat}).

For this partly conceptual study, we set $\mu_S=0$, as opposed to the more physical 
setting of strangeness neutrality $n_S=0$. (This more physical setting was
adopted in our recent work on contours of constant entropy in
Ref.~\cite{Borsanyi:2025dyp}.)

Similarly to our previous work in
Refs.~\cite{Fodor:2001au,Fodor:2001pe,Borsanyi:2022soo,Borsanyi:2023tdp,Borsanyi:2023wno,
Borsanyi:2024xrx} we used the reduced matrix formalism \cite{Hasenfratz:1991ax}
to obtain the configuration-specific expansion coefficients to arbitrary order.
This, together with our high statistics ensembles, allowed us to calculate Taylor
coefficients at $\mu_B=0$ to an unprecedented order $\mathcal{O}(\mu_B^{10})$.
At purely imaginary chemical potentials we calculated Taylor coefficients up to
order $\mathcal{O}(\mu_B^{4})$. Note that at nonzero imaginary chemical
potential the odd derivatives are also nonvanishing, in contrast to $\mu_B=0$
where only even derivatives appear in the Taylor expansion. Up to the sixth order in the Taylor expansion we observed
moderate cut-off dependence in this discretization scheme \cite{Borsanyi:2023wno}.

Our study of the Lee-Yang zeros is made possible by the impressive
dataset on baryon number fluctuations that the generated statistics has
allowed us to obtain. Thus, we start with the fluctuation results.

In Fig.~\ref{fig:chis} we show the $\mu_B$-derivatives of the pressure
\begin{equation}
\chi_n = \left( \frac{\partial^n (p/T^4)}{\partial (\mu_B/T)^n} \right)_{\mu_B=0}\rm,
\end{equation}
as functions of temperature on our ensembles. Due to our fixed volume setup,
we can calculate the sixth and eighth derivatives to unprecedented precision. In addition,
 for the first time in the literature, we provide a calculation of the 
tenth order coefficient as well.

There are several strategies for the computation of high order baryon fluctuations. The standard procedure involves the use of random sources
to compute the configuration-specific coefficients. These enter in connected and disconnected diagrams \cite{Allton:2005gk}. We, however,
follow Ref.~\cite{Borsanyi:2023wno} and use the exact representation
of the quark determinant as function of $\mu$, since for the given lattice size
this approach is more economic, and potentially more precise.
Even if only a single quark chemical potential is considered, the
number of terms in the expression for $\chi_2,\dots \chi_{10}$
is 2,8,25,79,230, respectively. Although we do not face a complex
action problem while we compute the $\chi_n$ coefficients, a remnant
cancellation problem appears through the cancellation between these terms,
contributing to the same $\chi_n$. We illustrate the severity of the problem
in Fig.~\ref{fig:cancellations}. For two lattice volumes we show the magnitude of the sum of all positive and all negative terms, respectively.
 These are constructed in such a way, that the actual signal -- namely the difference
between positive and negative contributions -- equals 1. With each successive
order the severity grows with a volume-dependent factor.
This plot explains why we refrain at present from the use of larger volumes,
and why more standard volumes, such as $32^3\times8$, are hopelessly unfeasible. 

We emphasize that for the data in Fig.~\ref{fig:chis} we did
not include input from imaginary-$\mu$ ensembles, thus the results do not rely on any sort of analytical assumption.\footnote{
We supply these coefficients as ancillary files along with the submission.
}

\subsection{Fluctuations at non-zero chemical potential}

Given the data in Fig.~\ref{fig:chis}, one can perform simple Taylor 
extrapolations of several different fluctuation observables to non-zero
$\mu_B$. For even derivatives we have:
\begin{align}
\chi^B_{2n}(T, \mu_B) &= \left( \frac{\partial^{2n} (p/T^4)}{\partial (\mu_B/T)^{2n}} \right)_{\mu_B} \\ &= \chi_{2n} + \chi_{2n+2} \frac{\hat \mu_B^2}{2!} + \chi_{2n+4} \frac{\hat \mu_B^4}{4!} + \dots \rm,
\end{align}
while for odd derivatives, we have:
\begin{align}
\chi^B_{2n+1}(T, \mu_B) &= \left( \frac{\partial^{2n+1} (p/T^4)}{\partial (\mu_B/T)^{2n+1}} \right)_{\mu_B} \\ &= \chi_{2n+2} \hat \mu_B + \chi_{2n+4} \frac{\hat \mu_B^3}{3!} + \dots \rm, 
\end{align}
where we have introduced the shorthand notation $\hat \mu_B = \frac{\mu_B}{T}$.
Note that we adopt the notation where 
\begin{equation*}
\chi_n(T) \equiv \chi^B_n(T, \mu_B=0)\rm.
\end{equation*}

\begin{figure}[t!]
    \centering
    \includegraphics[width=0.95\linewidth]{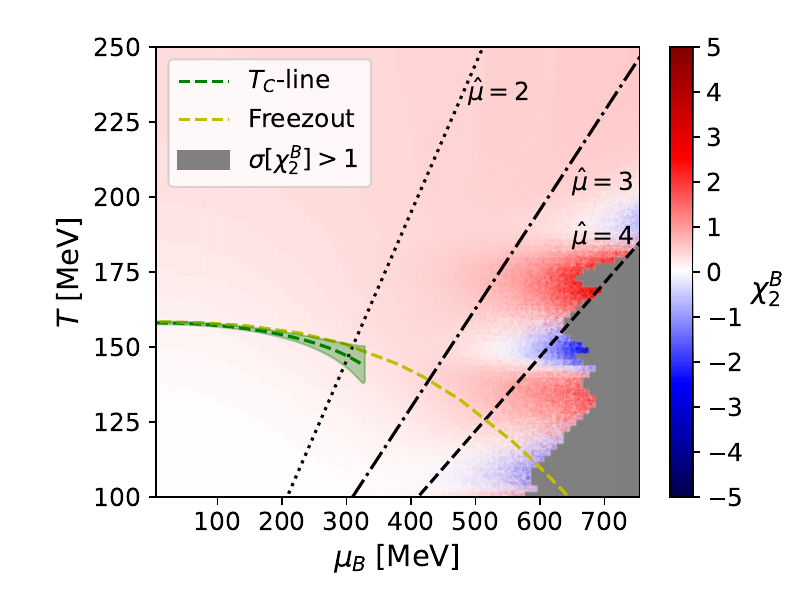}
    \caption{
    \label{fig:chi2_muB}
      $\chi^B_2(T, \mu_B)$ from a Taylor expansion to order $\chi_{10}$, using
      the spline Taylor expansion coefficients from Fig.~\ref{fig:chis}. 
      We also show the crossover line from Ref.~\cite{Borsanyi:2020fev} and the chemical 
      freezeout line from Ref.~\cite{Andronic:2005yp, Becattini:2012xb, Alba:2014eba, Vovchenko:2015idt}. 
      We added a noise to the color map to indicate the standard deviation of the observable. Additionally, the region where the magnitude of the absolute error exceeds $1$ is
      shown in gray. 
    }
    \end{figure}

We show $\chi^B_2(T,\mu_B)$ from a Taylor expansion up to order $\chi_{10}$
in Fig.~\ref{fig:chi2_muB}. At chemical potentials above $500$~MeV the estimate
shows the characteristic sign structure of $\chi_{10}$ from Fig.~\ref{fig:chis},
indicating that at such high chemical potentials, the highest Taylor
coefficient dominates the estimate. This means that without a further order in
the expansion, the correctness of this estimate 
at such high chemical
potentials cannot be tested. 

Ratios of fluctuations have also been studied in the literature.
In particular $\chi^B_4/\chi^B_2$ is interesting, as it is expected to 
show a non-monotonic behavior near the CEP, which can potentially be seen
quite far away from the CEP position itself~\cite{Stephanov:2011pb, Mroczek:2020rpm, Fu:2021oaw}. 
Another interesting ratio is  $\chi^B_3/\chi^B_1$, as it should 
change sign near the crossover in the vicinity of the CEP.

Although the Taylor expansion coefficients of ratios such as $\chi^B_3/\chi^B_1$ and
$\chi^B_4/\chi^B_2$ have been calculated in the literature before, this approach
is not suitable to reach high chemical potentials. This is because the 
radius of convergence of the Taylor expansion of the ratio
$\chi^B_4/\chi^B_2$ in the variable $\mu_B/T$ is close to $\pi/2$ at low
temperatures. This radius of convergence is unrelated to the critical endpoint, and is due to a zero of $\chi^B_2$ at purely imaginary chemical
potentials. This is explained in some details in Appendix A. 
Thus, this particular radius of convergence has nothing to do
with the CEP, but it makes a direct Taylor expansion of the ratio $\chi^B_4/\chi^B_2$
less useful.

\begin{figure*}[t]
    \centering
    \includegraphics[width=0.420\linewidth]{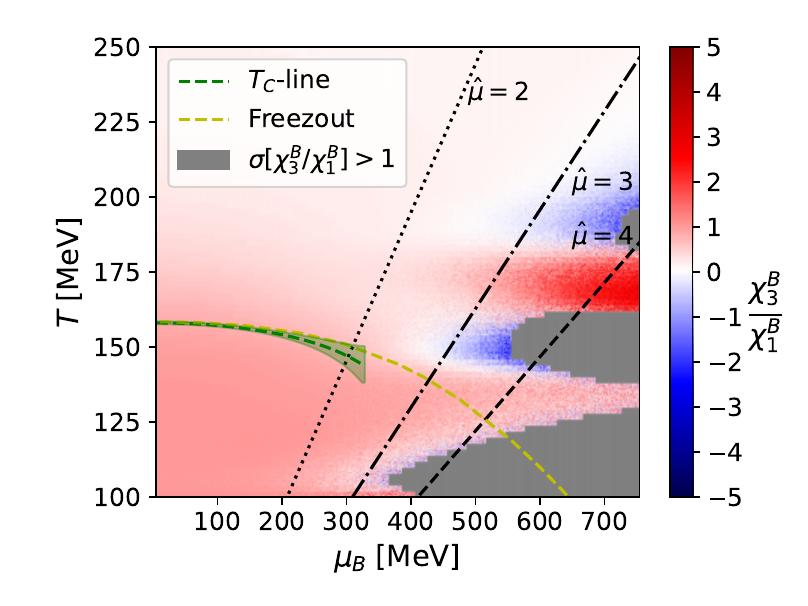}
    \includegraphics[width=0.420\linewidth]{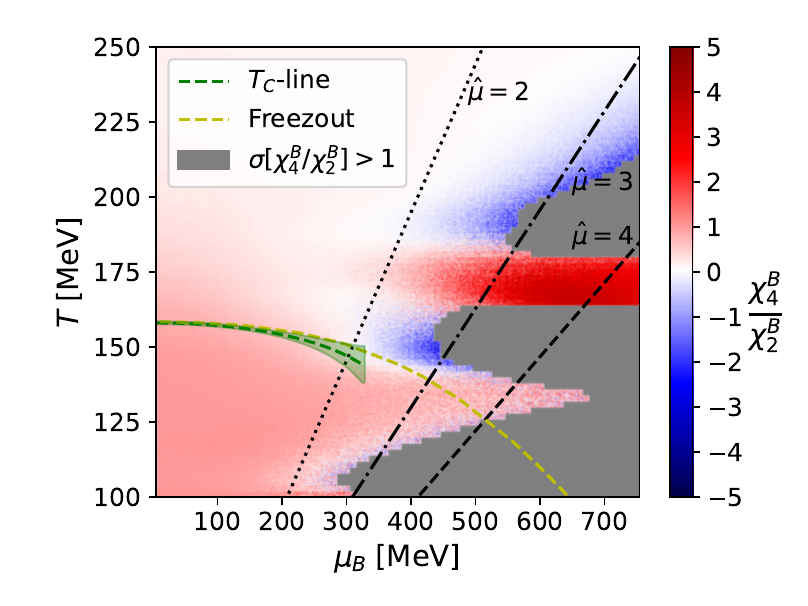}
    \caption{
    \label{fig:ratio_muB}
      $\chi^B_3(T, \mu_B)/\chi^B_1(T, \mu_B)$ (left) and 
      $\chi^B_4(T, \mu_B)/\chi^B_2(T, \mu_B)$ (right) estimated using
      Eq.~\eqref{eq:ratio42} and the splined Taylor expansion coefficients from Fig.~\ref{fig:chis}. We added a noise to the color map to indicate the standard deviation of the observable. Additionally, the region where the magnitude of the absolute error exceeds $1$ is
      shown in gray. 
    }
\end{figure*}

A safer approach is, thus, to Taylor expand the numerator and denominator separately:
\begin{equation}
\label{eq:ratio42}
\frac{\chi^B_4(T,\mu_B)}{\chi^B_2(T,\mu_B)} \approx 
\frac{
\chi_4  + \chi_6 \frac{\hat \mu_B^2}{2}
+ \chi_8 \frac{\hat \mu_4^6}{4!}
+ \chi_{10} \frac{\hat \mu_6^8}{6!} \rm,
}{\chi_2 + \chi_4 \frac{\hat \mu_B^2}{2} + \chi_6 \frac{\hat \mu_B^4}{4!}
+ \chi_8 \frac{\hat \mu_B^6}{6!}
+ \chi_{10} \frac{\hat \mu_B^8}{8!}
}
\end{equation}
and similarly for $\chi^B_3 / \chi^B_1$. The results of this 
approximation are shown in Fig.~\ref{fig:ratio_muB}. In both Figures,
we see a similar structure as in $\chi^B_2$ itself: at large chemical
potentials, the highest order of expansion dominates.
At this level of truncation the ratio $\chi^B_4/\chi^B_2$ does show non-monotonic behavior, though the $\chi^B_3/\chi^B_1$ ratio does not change sign 
near the transition and freeze-out lines. Thus, the picture is not completely consistent
with the expected signatures of chiral criticality. Rather, the source of the non monotonic behavior of $\chi^B_4/\chi^B_2$ is more likely
to be the truncation of the Taylor series. 
We are directly facing the limitations of what is possible for our current dataset with Taylor extrapolation.

In order to have a more quantitative look at the phase diagram, we now turn
to the discussion of Lee-Yang zeros.

\section{Lee-yang zeros from rational function approximations}

\subsection{Lee-Yang zeros}
Lee-Yang zeros are zeros of the partition function in the complex chemical potential plane. For the case of a first-order phase transition they approach the real line at
an infinite volume as $\operatorname{Im} \mu_B \sim V^{-1}$, where $V$ is the physical
volume. In contrast, for a second order phase transition the zeros approach
the real axis as $\operatorname{Im} \mu_B \sim V^{-\frac{\gamma/\nu+D}{2D}}$,
where $\gamma$ and $\nu$ are the usual critical exponents and $D$ is the
dimensionality of space~\cite{Itzykson:1983gb}. For 3D Ising critial exponents
this corresponds to $\operatorname{Im} \mu_B \sim V^{-0.83}$. In both the first
and second order cases, an infinite number of zeros accumulates near the real
line, leading to the singularity (phase transition) in the thermodynamic limit.
For a crossover, there is no general formula for the volume scaling of the
Lee-Yang zero positions, except for the case when the crossover is in the
immediate vicinity of a critical point. In this case, the finite volume scaling
is of the form~\cite{Itzykson:1983gb}  $\operatorname{Im} \mu_B \sim A + B
V^{-\frac{1}{1 + \sigma}}$, where $\sigma$ is the so-called edge exponent. Its
most common value~\cite{Butera:2012tq,Gliozzi:2013ysa} is $\sigma\approx0.076$, leading
to the behavior:
\begin{equation}
\operatorname{Im} \mu_B \sim A + B V^{-0.93} \rm,
\end{equation}
where $A$ and $B$ are non-universal amplitudes. Since we only have one volume
in our study, we are going to assume that the second term is negligible $A \gg B V^{-0.93}$. 
The temperature dependence of $A$ is given by the analytic continuation
of the Kadanoff scaling ansatz to be:
\begin{equation}
\label{eq:KadanoffAnaCont}
\operatorname{Im} \mu_B \sim A \sim |T-T_{CEP}|^{\beta \delta}, 
\end{equation}
where for the 3D Ising universality class the product of critical exponents
$\beta \delta \approx 1.56$. By knowing $\operatorname{Im} \mu_B$ for several
temperatures above the temperature of the critical endpoint,
Eq.~\eqref{eq:KadanoffAnaCont} can be used to fit the 
temperature of the critical endpoint $T_{CEP}$. This basic approach, also used
in Refs.~\cite{Basar:2023nkp, Clarke:2024ugt}, is the one we pursue here.

In order to calculate such an estimate
of $T_{CEP}$, one first needs to estimate
the Lee-Yang zero positions at several fixed values of the temperature $T$. How
this is achieved is discussed in the next subsection.

Here, we note that Eq.~\eqref{eq:KadanoffAnaCont} is only valid in the close vicinity
of the CEP. However, extracting the position of the Lee-Yang zeros at temperatures
well below the crossover has so far proven to be prohibitively difficult. Thus,
current extrapolations only use data in the vicinity of the crossover
temperature (just below $160$MeV). This means that one is likely using data with sizable
corrections to the scaling ansatz of 
Eq.~\eqref{eq:KadanoffAnaCont}. We will come back to this issue later, when we discuss the systematic
errors on the CEP estimate.

\subsection{Rational approximations and the choice of an expansion variable}

To look for the Lee-Yang zeros, one needs a functional form for the baryon
number susceptibilities that allows for the existence of poles. The simplest
option is a rational function: the ratio of polynomials.

The direct lattice data from the simulations consist of Taylor expansion
coefficients in $\mu_B^2$ (see Fig.~\ref{fig:chis}) at $\mu_B=0$ (or in $\mu_B$
at non-zero imaginary $\mu_B$). Only even terms appear in the Taylor expansion
because of charge conjugation symmetry. Converting this Taylor series to a
rational approximation leads to an ansatz of the form:
\begin{equation}
\label{eq:Pade}
F(T,\mu_B) = \frac{p_0 + p_1 \hat \mu_B^2 + \dots + p_m \hat \mu_B^{2m}}{1 + q_1 \hat \mu_B^2 + \dots + q_n \hat \mu_B^{2n}} \rm,
\end{equation}
where $F$ is some physical quantity of interest, such as $\chi^B_2$.
Such an ansatz was used in Ref.~\cite{Clarke:2024ugt}, where
odd terms in $\mu_B$ were also allowed.

In general, one can use the chain rule to convert these expansion coefficients
to expansion coefficients in different variables. The variable we will consider in this
work is 
\begin{equation}
x=\cosh\left(\hat \mu_B\right)-1=\frac{1}{2}\hat \mu_B^2 + \mathcal{O}(\hat \mu_B^4) \rm.
\end{equation}
An expansion in this variable has two benefits. First, in the hadronic phase 
(below the crossover temperature) we expect this expansion to converge faster 
than a Taylor expansion: in the Boltzmann approximation of the
ideal hadron resonance gas model, only the 
leading order term is non-zero. While this first term expresses the baryon density,
the next order is responsible for the repulsive baryon interactions, and is
suppressed by several orders in magnitude
\cite{Vovchenko:2017xad, Huovinen:2017ogf, Bellwied:2021nrt}. Another way
to phrase this is: in this expansion variable, the leading order term is
dominated by single baryons, while the next-to-leading order term is only
non-zero in case of genuine two-baryon correlations. 
The second advantage of
this variable is
that it makes manifest use of the Roberge-Weiss symmetry: the periodicity of the QCD
free energy as a function of imaginary chemical potential~\cite{Roberge:1986mm}.
This leads to the ansatz:
\begin{equation}
\label{eq:CoshPade}
F(T,\mu_B) = \frac{p_0 + p_1 x + \dots + p_m x^{m}}{1 + q_1 x + \dots + q_n x^{n}} \rm.
\end{equation}
Ansätze of this form will be considered in this paper for the first time in the
literature. We, however, will not consider ans\"atze of the form 
of Eq.~\eqref{eq:Pade}, since by not respecting Roberge-Weiss symmetry, they are unreliable for $|\mu_B/T|>\pi$ (at least in some directions in the complex $\mu_B$ plane). Specifically, we will use $m=1$ and $n=2$. This specific choice is motivated by three observations. First, the $n=2$ is the lowest order in the denominator that allows for an arbitrary position of the Lee-Yang zeros in the strip of the complex plane with $-\pi<\operatorname{Im} \mu_B\leq\pi$ (the strip is then periodically repeated). Second, increasing the number of parameters, e.g. taking $n=3$, we observed an overfitting of the data, and a corresponding instability in the Lee-Yang zero positions. Third, using fewer parameters (so $m=0$ instead of $m=1$) did not allow us to fit all the data with good enough $\chi^2$ values.
To have a systematic error estimate
on the Lee-Yang zero extraction from
the Padé approximants, we will consider, however, different choices for the observable $F$.

\begin{figure*}
    \includegraphics[width=0.49\textwidth]{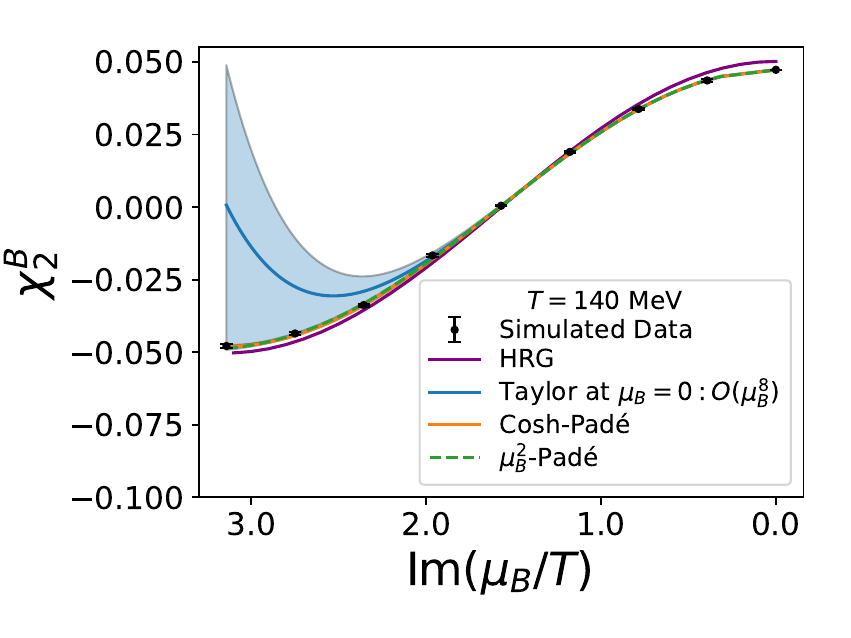}
    \includegraphics[width=0.49\textwidth]{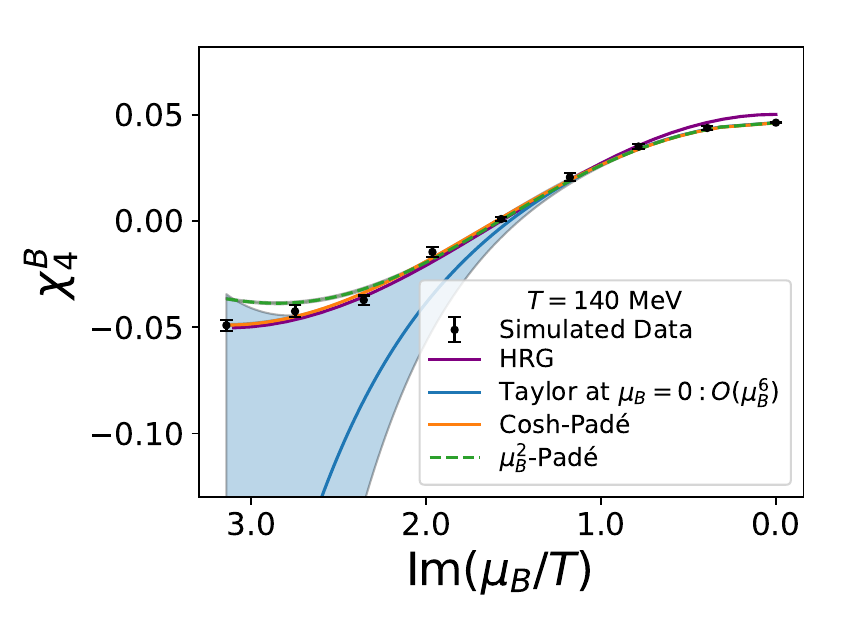}
    \caption{\label{fig:coshpade}
    The result of the fitting of the model in Eq.~\eqref{eq:CoshPade} with
    $m=1$ and $n=2$, using $x=\cosh(\mu_B/T)$ (orange) or $x=\mu_B^2$ (green). 
    The less sophisticated ansatz with $x=\mu_B^2$ offers a poor fit quality, whereas the periodic one is always acceptable.
    This example shows the fit at $T=140$~MeV  with the $\Delta \hat p$ ansatz. Different temperatures are fitted independently.
    }
\end{figure*}

We show the resulting rational function in Fig.~\ref{fig:coshpade} in the range where we have data points ([$\pi$,0] in imaginary $\mu_B/T$).
We picked the example of $T=140$~MeV with the $\Delta \hat p$ ansatz and selected $\chi_2$ and $\chi_4$ as observables. The fit includes $\chi_1,\dots,\chi_4$ at 8 imaginary valued chemical potentials, and $\chi_2,\dots,\chi_{10}$ at $\mu_B=0$. The shown example has a reduced chi square $\chi^2/\mathrm{ndof}=45.5/30$.

Note that if one has further information about the analytic structure of the
free energy, it is possible to use holomorphic maps to take advantage of this information.
In particular Ref.~\cite{Basar:2023nkp} used a method that moves the cut formed
from the accumulation of Lee-Yang zeros in the thermodynamic limit (the Lee-Yang
edge) to the unit circle.  When Ref.~\cite{Basar:2023nkp} applied this approach
to the then available lattice data, it found that it made only a small
difference compared to the naive Padé approximation approach of Eq.~\eqref{eq:Pade}
in the final results. We will see that other effects lead to a larger
systematic error in the Lee-Yang zero positions, and thus we will not apply this
approach in our study. In addition to this basic observation, one should also
note that Ref.~\cite{Basar:2023nkp} assumes that cuts starting at the Lee-Yang
edge are the only non-analyticities in the free energy. However, in full QCD,
this is not correct, as the Roberge-Weiss periodicity means that such
discontinuities are repeated an infinite number of times in the imaginary
chemical potential direction. How these affect the singularity structure after
applying such a holomorphic map has not yet been worked out and should be the
subject of future theoretical research.

\subsection{The hadron resonance gas baseline}

A common non-critical baseline for the critical endpoint search in QCD phenomenology
is the hadron resonance gas (HRG) model. In terms of the
Taylor coefficients, it satisfies:
\begin{equation}
\label{eq:HRGequality}
\chi_2 = 
\chi_4 = 
\chi_6 = 
\chi_8 = \dots \rm.
\end{equation}
Strictly speaking this relation is true in the Boltzmann approximation only, which however is well justified for baryons.
While the leading $\chi_2$ coefficient
as a function of temperature depends on the actual baryon spectrum, the prediction of Eq.~\eqref{eq:HRGequality} is independent of the exact hadron
spectrum used in the construction of the model.
Thus, at a finite lattice spacing (where the hadron
spectrum is distorted by cut-off effects) Eq.~\eqref{eq:HRGequality} still remains true. This also holds in a finite volume, such as those in our simulations.

The HRG features no critical endpoint. Indeed, by applying the 
ansatz of Eq.~\eqref{eq:CoshPade} one should get $q_1=...=q_n=0$, since the pressure in the HRG is simply $p = p_0(T) + p_1(T) \cosh(\mu_B/T)$. By including noise, non-zero coefficients will appear in the denominator. However, in general the position of the zeros of the denominator will not be stable under variations in the noise itself.

In fact, the zeros of the denominator:
\begin{equation}
\begin{aligned}
1 + q_1 (\cosh(\hat \mu_B)-1) + q_2(\cosh(\hat \mu_B)-1)^2
\end{aligned}
\label{eq:denom}
\end{equation}
are bound to be either close to $\Im \mu_B= \pi/2$ or close to
$\mathrm{Im}~\mu_B=\pi$, depending on the particular realization of the
noise. This is because the imaginary part of the solution to Eq.~\eqref{eq:denom} changes abruptly from $\pi$ to around $\pi/2$ as the sign of the $q_2$ coefficient changes, as discussed in more detail in Appendix B. This leads to an instability in the noisy HRG model, where a Lee-Yang zero with either $\operatorname{Im} \mu_B \approx \pi$ or 
$\operatorname{Im} \mu_B \approx \pi/2$ can appear. 

This expectation is confirmed by a mock data analysis, where we used noise with the same relative errors and correlations as our lattice data in addition to the HRG
prediction for the Taylor coefficients, and performed the same analysis we employed with the actual lattice data. When we show noisy HRG baselines in this paper, we always show both branches of this instability.

A lattice result consistent with such a noisy HRG
baseline does not rule out a CEP. However,
it may indicate that we have reached the 
limits of the applicability of our ansatz.
Baryon interactions are encoded in the higher orders ($k\ge2$) of the fugacity expansion 
$p/T^4 = \sum_k c_k \cosh(k \hat \mu_B)$. We know that $c_2, c_3, \dots$ are suppressed by Boltzmann
factors related to two-, three-baryon states, respectively, at low $T$~\cite{Vovchenko:2017xad,
Huovinen:2017ogf, Bellwied:2021nrt}. No matter 
how small the sub-leading corrections are,
they appear in the higher order 
baryon number fluctuation $\chi_{2n}$ enhanced by a factor $k^{2n}$,
and will give a dominant contribution for a high 
enough order of the Taylor coefficients.
Thus, by taking a sufficiently large number of 
derivatives, one always expects
deviations from the HRG model.
By not observing such deviations, we see that either
the order of our Taylor expansion is not high enough or
the statistics is not sufficient (thus, the uncertainties are not small enough)
to resolve deviations from the HRG model and thus to constrain the analytic
structure of QCD at the given temperature. 
Thus, we will always
compare our Padé approximant's poles with the HRG
baseline. Whenever we find agreement, we will
exclude those low temperatures from our fit range for
the scaling ansatz in Eq.~\eqref{eq:KadanoffAnaCont}.
For our dataset, this means that we will have to restrict our
fits of the scaling ansatz to temperatures above $130$~MeV. 
Another reason to exclude these data points from further analysis
is that they share
the same instability in the imaginary part of the Lee-Yang zeros, which fluctuates between $\pi$ and approximately $\pi/2$.

\begin{figure}[t]
    \centering
    \includegraphics[width=0.850\linewidth]{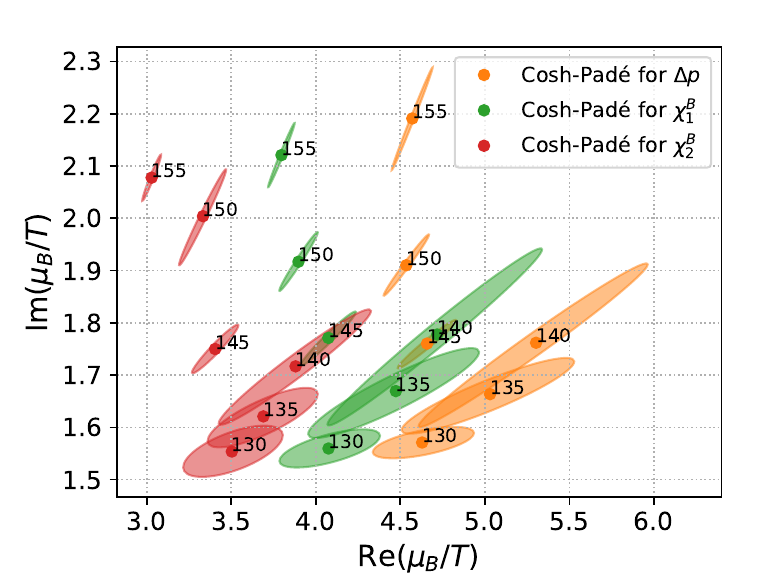}
    \caption{
    \label{fig:systematics_LY}
    The extracted Lee-Yang zeros positions for various
    temperatures in the complex $\mu_B/T$ plane. 
    The ellipses represent statistical errors.
    The different colors represent the choices $F=\Delta p$ (orange), $F=\chi^B_1$ (green) and $F=\chi^B_2$ (red) respectively. The temperatures (in MeV) are indicated next to the particular error ellipse.
}
\end{figure}

\subsection{Systematics of the Lee-Yang zero extraction}

One has to remember that even before the scaling ansatz in Eq.~\eqref{eq:KadanoffAnaCont} is used, each data point on the position of the leading Lee-Yang zero at 
the different fixed values of the temperature is based on an analytic
continuation to complex chemical potentials. Such an
analytic continuation could have significant 
systematic errors. To estimate these, we use three
different ans\"atze for the free energy, distinguished
by a different choice of $F$ in Eq.~\eqref{eq:CoshPade}.
We choose either $\Delta \hat p (T, \hat \mu_B) = \hat p(T,\hat \mu_B) - \hat p(T,0)$, $\chi^B_1$ or $\chi^B_2$, where 
$\hat p = p/T^4$.
The impact of these three different choices is illustrated in Fig.~\ref{fig:systematics_LY}, where the estimated Lee-Yang zero positions are shown for different temperatures
with the three different choices. A conclusion one can draw from this figure is that, while the imaginary part of the Lee-Yang zeros is stable for different choices of $F$, the real part is not.

\subsection{Systematics of the extrapolation to the CEP}

\begin{figure*}[t]
    \centering
    \includegraphics[width=0.420\linewidth]{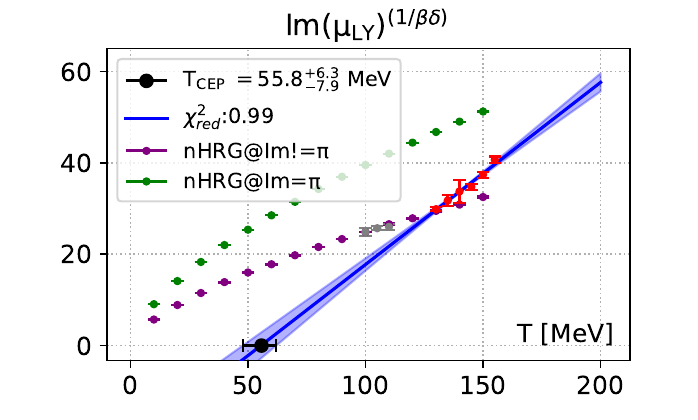}
    \includegraphics[width=0.420\linewidth]{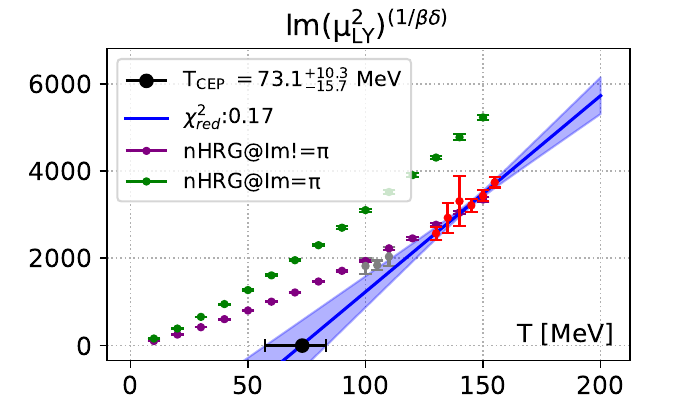}
    \includegraphics[width=0.420\linewidth]{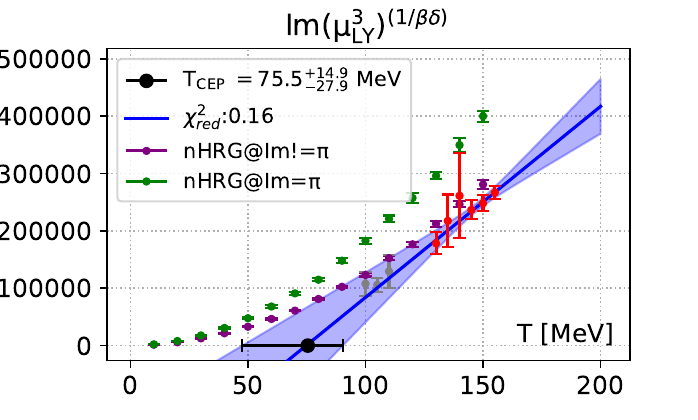}
    \includegraphics[width=0.420\linewidth]{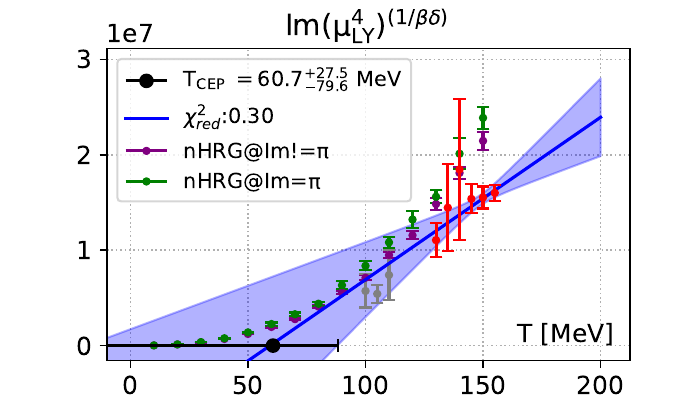}

    \caption{
    \label{fig:extra_var}
    Extrapolation of the critical temperature using the
    four different ansätze in Eq.~\eqref{eq:scaling_variables}. 
    Near the CEP, all four ansätze are asymptotically 
    equivalent, though farther away they lead to different
    behavior. All four fits
    have good $\chi^2$ values, but they lead to different values for the critical temperature. The two (unstable) branches of the noisy HRG model (nHRG) baseline is also shown.
}
\end{figure*}

Supposing that the CEP exists, universality arguments give information about the asymptotic behavior of thermodynamic functions in the
vicinity of the critical endpoint. The asymptotic behavior should be of the form given by Eq.~\eqref{eq:KadanoffAnaCont}. 
However, at the moment we only have
data far away from the critical endpoint, where the
Lee-Yang zeros are still far from the real axis. We 
have to perform an extrapolation to a rather distant 
temperature, to get an estimate of the critical endpoint.
To estimate the systematic errors of such a bold extrapolation, we performed fits using four 
different ans\"atze. All of which have the same 
asymptotic behavior, but differ away from the critical
endpoint:
\begin{equation}
\label{eq:scaling_variables}
\begin{aligned}
T-T_{CEP} \approx A \left( \operatorname{Im} \mu_{LY} \right)^{\frac{1}{\beta \delta}} \rm, \\
T-T_{CEP} \approx B \left( \operatorname{Im} \mu_{LY}^2  \right)^{\frac{1}{\beta \delta}} \rm, \\ 
T-T_{CEP} \approx C \left( \operatorname{Im} \mu_{LY}^3 \right)^{\frac{1}{\beta \delta}} \rm, \\
T-T_{CEP} \approx D \left( \operatorname{Im} \mu_{LY}^4 \right)^{\frac{1}{\beta \delta}} \rm. \\ 
\end{aligned}
\end{equation}
In the vicinity of the CEP we have 
\begin{equation*}
\begin{aligned}
\operatorname{Im} \mu_{LY}^2 \sim 2 \mu_{CEP} \operatorname{Im} \mu_{LY}, \\
\operatorname{Im} \mu_{LY}^3 \sim 3 \mu_{CEP}^2 \operatorname{Im} \mu_{LY} -(\operatorname{Im} \mu_{LY})^3, \\
\operatorname{Im} \mu_{LY}^4 \sim  4 \mu_{CEP}^3 \operatorname{Im} \mu_{LY} - 4 \mu_{CEP} (\operatorname{Im} \mu_{LY})^3 \rm.
\end{aligned}
\end{equation*}
Since all of these functions have the same asymptotic behavior near the critical endpoint, there is no reason to prefer
one of these ans\"atze to the others. In Fig.~\ref{fig:extra_var} we show these 
four different fits on the same data, together with the noisy HRG baseline for comparison. Below $T=130$~MeV we find that the lattice results are
consistent with the baseline. As explained previously,
this means that the data quality in this range of temperatures
is not sufficient to constrain the analytic structure of the
QCD free energy. Thus, those temperatures (shown in grey in Fig.~\ref{fig:extra_var}) are left out from our fits.
With this temperature range restriction, one can see that depending on the choice
of the scaling variable, one can get a
CEP prediction compatible with the nuclear liquid-gas transition, and just as well one compatible with a high-temperature chiral critical endpoint.

\begin{figure}[t]
    \centering
    \includegraphics[width=0.850\linewidth]{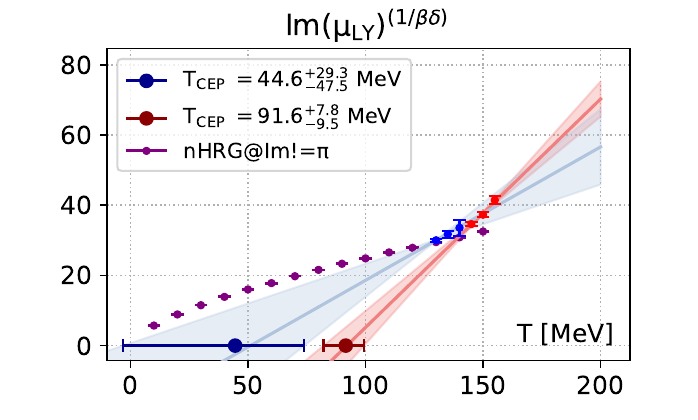}
    \caption{
    \label{fig:extra_range}Extrapolation of $T_{CEP}$ using two different temperature ranges.
}
\end{figure}

Similarly important is the choice of the
temperature range of the fit. This is illustrated in Fig.~\ref{fig:extra_range}., where one can see that the systematic error due to the choice of the temperature range is also rather large.

Let us also underline an important difference between extrapolations based on Eq.~\eqref{eq:KadanoffAnaCont} and other 
extrapolations in lattice field theory, such as a continuum limit extrapolation. 
In our case each point (each temperature) used for the
extrapolation is itself the result of an extrapolation in the complex chemical
potential plane.  The lower the temperature, the more relevant
the data point is , due to its closer proximity to the critical end point.
Thus, although data at lower temperatures more reliably guide the extrapolation toward a Lee-Yang zero
on the real $\mu_B$ axis, they require much higher statistics to combat the still unresolved sign problem.
This is because the higher orders in the extrapolation
scheme are the ones that will be most sensitive to baryon interactions, but also the most difficult ones to determine. In short, 
unfortunately, the reliability of the data diminishes with decreasing temperature.
This is in contrast with the continuum extrapolation, where at smaller lattice
spacing the simulations are only more expensive, but systematic errors do not
rise. While fine lattices are always included in the continuum extrapolations,
in our systematic analysis we will also consider extrapolations where some of the lowest temperatures (closer to a CEP) are discarded.

\begin{figure*}[t]
    \centering
    \includegraphics[width=0.420\linewidth]{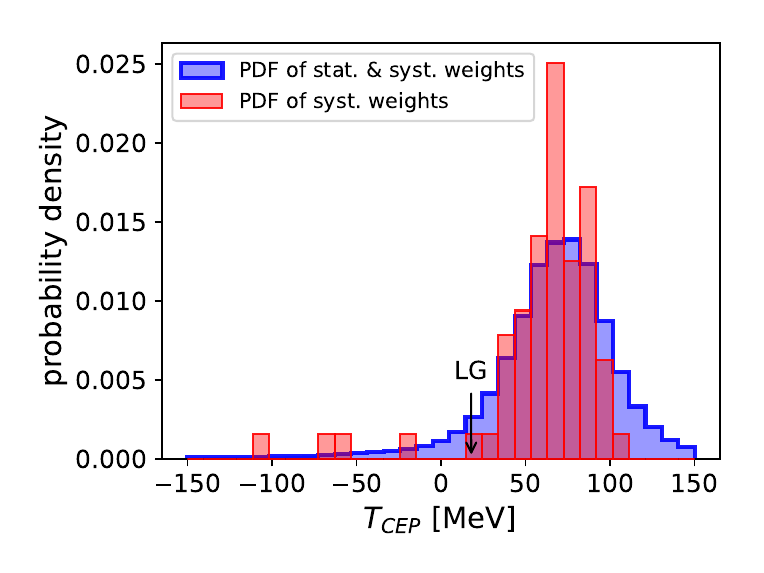}
    \includegraphics[width=0.420\linewidth]{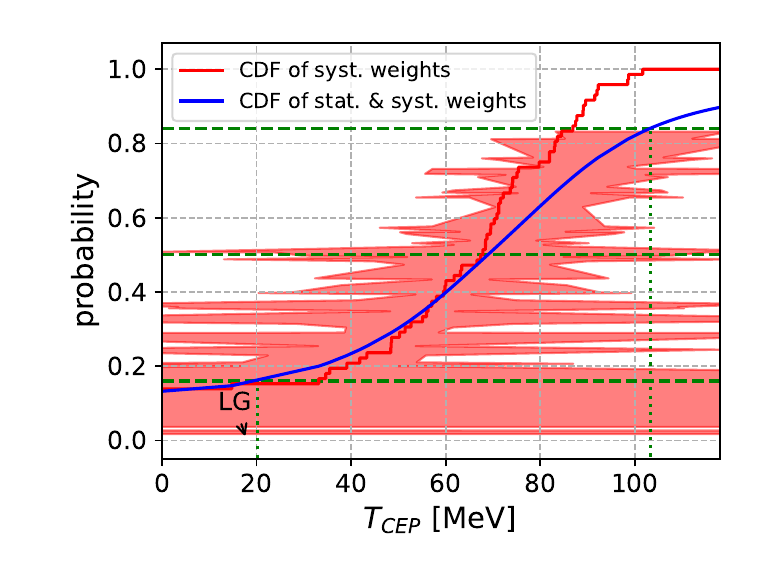}
    \caption{
    \label{fig:histogram_final}
    The histogram (left) and the cumulative distribution function for the
    critical endpoint temperature $T_{CEP}$. (We indicate the location of the liquid-gas critical endpoint with the label LG \cite{Elliott:2013pna}).
}
\end{figure*}

\subsection{Weighting and posterior distribution for $T_{\rm{CEP}}$}

In our final results we will consider the following choices to estimate systematic errors:
\begin{enumerate}
\item The different choices for the analytic continuation to complex chemical potentials, corresponding to $F=\Delta p,$ $\chi^B_1$ or $\chi^B_2$ 
respectively in Eq.~\eqref{eq:CoshPade} (3 choices).
\item Four different choices of the scaling ansatz to extrapolate the position of the critical endpoint, shown in Eq.~\eqref{eq:scaling_variables}.
\item Six different choices for the temperature range in the extrapolations in Eq.~\eqref{eq:scaling_variables}.
The choices are: \\
$[130\rm{MeV}, 145\rm{MeV}]$, 
$[135\rm{MeV}, 150\rm{MeV}]$, \\
$[140\rm{MeV}, 155\rm{MeV}]$, 
$[130\rm{MeV}, 150\rm{MeV}]$,  \\
$[135\rm{MeV}, 155\rm{MeV}]$, 
$[130\rm{MeV}, 155\rm{MeV}]$. \\

\end{enumerate}
These lead to a total of $3 \times 4 \times 6 = 72$ different analyses. All of the fits have acceptable $\chi^2$ values, thus we combine them with uniform weights. The different analyses are combined into a single distribution function for $T_{CEP}$ by assuming Gaussian distributions 
for the individual fits. Combining the different fits into a single cumulative distribution function, we then proceed using the following formula:
\begin{equation}
CDF(T_{CEP}) = \sum_j w_j \frac{1}{2} \left( 1- \operatorname{erf} \left( \frac{m_j -T_{CEP}}{\sqrt{2 \sigma_j}} \right) \right) \rm,
\end{equation}
where $m_j$ and $\sigma_j$ are the mean and the statistical error of the given 
analysis. Since we use uniform weights, we take $w_j=1/72$ for all fits. 

Taking the results at face value, we arrive at an upper bound on
the location of the chiral critical endpoint position. With
$84\%$ probability, the critical endpoint is either
below $T=103$~MeV or it does not exist.
Assuming that the CEP exists, its most likely position (the peak
of the posterior distribution for $T_{\rm{CEP}}$)
is in the temperature bin near $70$~MeV.

The probability distribution for $T_{CEP}$ has a non-negligible  
part in the very low temperature range. This means that the data is consistent with the non-existence of chiral 
critical endpoint, while the Lee-Yang zeros getting
closer to the real axis being a consequence of the existence 
of the nuclear liquid-gas critical endpoint. Such a scenario
was presented e.g. in Ref.~\cite{Vovchenko:2016rkn}, where in a van der Waals type
generalization of the hadron resonance gas model, echoes of the
nuclear liquid-gas critical endpoint could be observed at surprisingly
large temperatures.

\section{Summary and discussion}

We have performed large statistics lattice
simulations using 4HEX improved staggered
fermions on $16^3 \times 8$ lattices at
zero and purely imaginary 
baryochemical potentials. On
these ensembles, we calculated 
Taylor coefficients of the pressure up to
tenth order at zero, and up to the fourth order
at imaginary chemical potentials. We used
this data to construct rational function approximations of the QCD free energy, and
to approximate the position of the Lee-Yang zero closest to the origin in the
complex chemical potential plane. The 
imaginary part of the Lee-Yang zeros was then
extrapolated to zero value, in order to estimate the temperature of the critical
endpoint of QCD.

The main result of our manuscipt is an upper bound on
the location of the critical endpoint position. With
$84\%$ probability, the critical endpoint is either
below $T=103$~MeV or it does not exist.
Assuming that the CEP exists, its most likely position is near $70$~MeV.

Our results are 
compatible with the estimates of the CEP position from 
functional methods~\cite{Isserstedt:2019pgx, Gao:2020qsj}, but are in slight tension with recent
lattice estimates~\cite{Basar:2023nkp, Clarke:2024ugt}.
Note that neither these, nor our results have been continuum extrapolated, and the lattice artefacts are not expected to be the same.

Our work has some important technical caveats:
\begin{enumerate}
\item The continuum limit was not performed, as only $N_\tau=8$ lattices were used.
\item The thermodynamic limit was not taken, nor a finite volume scaling analysis was performed.
\item Because of a finite order truncation of the Taylor expansions, the Lee-Yang zero extractions at fixed temperature suffer from systematic errors.
\item The extrapolation of the critical temperature $T_{\rm{CEP}}$ also suffers from  systematic errors, since the Lee-Yang zero positions have to be extrapolated to a region of the phase diagram that is quite far away from where the lattice data is available.
\end{enumerate}

In principle it is possible to alleviate all four of these problems,
but for the moment they lead to a rather large uncertainty in 
the position of the CEP.

Caveats 1 and 2 cannot be addressed with the dataset used for this work. 
While in our recent publication~\cite{Borsanyi:2023wno} we observed quite mild cut-off effects for this discretization for Taylor coefficients up to sixth order, this does not automatically
guarantee similarly small cut-off effects for the 8th and 10th
order coefficients calculated in this work. Similarly, 
for temperatures below the crossover, our earlier comparison
between the continnum extrapolated data~\cite{Borsanyi:2023wno} at a smaller volume $LT=2$ and
our $N_t=12$ 4stout improved data at a larger volume ($LT=4$) 
were compatible within errors, indicating mild finite volume 
effects. However, this does not guarantee that the same
would hold for the larger 8th and 10th order coefficients calculated
in this work. Ultimately one has to perform a similar analysis
for at least two larger volumes and at least two smaller lattice
spacings. In view of Fig.~\ref{fig:cancellations}, this will be a serious challenge, since the costs of the eighth order coefficient scales with the sixth power of the volume.

On the other hand, caveats 3 and 4 are considered in our final systematic errors, and lead to the facts that our analysis
does not provide a sharp prediction for the CEP position,
but only a bound. Previous works in Refs.~\cite{Clarke:2024ugt, Basar:2023nkp} have not
taken all of these systematic effects into account. This is the reason
why we obtain significantly larger final errors, even though 
our input data are much more precise.
Caveat 3 is the main reason why only the temperature of the critical endpoint can be estimated, but not its chemical potential. The real part of the position of the Lee-Yang
zeros is much less stable under a change of the anasatz used for the analytic continuation compared to the imaginary part. 

In principle, it is also possible to extract Lee-Yang zeros
without relying on Taylor coefficients, and thus without any truncation errors, using the canonical
formalism as was done in Ref.~\cite{Giordano:2019gev} for $N_\tau=4$ lattices. This method has a good chance to
significantly reduce or even completely remove issue 3.
How the signal-to-noise ratios of this method change at finer lattices, however, remains to be seen.

Extending the temperature range of the available lattice
simulations should help with caveat 4. However, this is not
a trivial task, as the signal to noise ratio of the Taylor 
coefficients also deteriorates at smaller temperatures. In addition, corrections to the hadron resonance gas model (the subleading
fugacity expansion coefficients) become smaller at low
temperatures, making the contributions responsible for criticality
harder to access. Thus, extending datasets to lower temperatures
might require new algorithmic ideas.
Of course, as the quality of the data increases and deviations from the hadron resonance gas can be resolved at
lower temperatures, one must also increase the number of free parameters
in the ansatz used to extract the Lee-Yang zeros. 

\subsection{Acknowledgements}
This work is supported by the MKW NRW under the funding code NW21-024-A. 
Further funding was received from the DFG under the Project No. 496127839. 
This work was also supported by the Hungarian National Research, Development and Innovation
Office, NKFIH Grant No. KKP126769.
This work was also supported by the NKFIH excellence grant TKP2021{\textunderscore}NKTA{\textunderscore}64.
This work is also supported by the Hungarian National Research,
Development and Innovation Office under Project No. FK 147164.
This material is also based upon work supported by the National Science Foundation under grants No. PHY-2208724,
and PHY-2116686, and within the framework
of the MUSES collaboration, under grant number No. OAC-
2103680. This material is also based upon work supported
by the U.S. Department of Energy, Office of Science, Office of Nuclear Physics, under Award Numbers DE-SC0022023 
and DE-SC0025025, as well as by the National Aeronautics and Space Agency (NASA)  under Award Number 80NSSC24K0767. 
The authors gratefully acknowledge the Gauss Centre for
Supercomputing e.V. (\url{www.gauss-centre.eu}) for funding
this project by providing computing time on the GCS
Supercomputer HAWK at HLRS, Stuttgart.
An award of computer time was provided by the INCITE program. This research used resources of the Argonne Leadership Computing Facility, which is a DOE Office of Science User Facility supported under Contract DE-AC02-06CH11357.

\appendix

\section{Zeros of $\chi^B_2$ and $\chi^B_4$ at purely imaginary baryochemical potential}
The simplest way to understand why both $\chi^B_2$ and $\chi^B_4$ have a zero
at purely imaginary chemical potential at temperatures below the crossover 
is to first start with the hadron resonance gas model (in the Bolzmann 
approximation), where the pressure is approximated to be:
\begin{equation}
p(T, \mu_B) \approx p_0(T) + p_1(T) \cosh\left( \frac{\mu_B}{T}\right) \rm.
\end{equation}
Differentiating with respect to $\mu_B/T$ and writing $\mu_B=i \mu_I$ we get a zero
for both $\chi^B_2$ and $\chi^B_4$ at $\mu_I/T=\pi/2$. This zero is a property of 
a free gas of hadrons. The ratio $\chi^B_4 / \chi^B_2$ is equal to $1$ in this model.
Near $\mu_I=\pi/2$ this value of $1$ is realized as a $0/0$ limit.
However, the HRG model
is not exact. A more accurate approximation is given by the next order of the
fugacity expansion~\cite{Vovchenko:2017xad, Huovinen:2017ogf, Bellwied:2021nrt}:
\begin{equation}
p(T, \mu_B) \approx p_0(T) + p_1(T) \cosh\left( \frac{\mu_B}{T}\right) + p_2(T) \cosh\left( 2 \frac{\mu_B}{T}\right) \rm,
\end{equation}
where $p_2(T) \ll p_1(T)$. Now differentiating with respect to $\mu_B/T$ 
and writing $\mu_B=i \mu_I$ we arrive at the conclusion that 
both $\chi^B_2$ and $\chi^B_4$ have a zero in the vicinity of 
$\mu_I/T=\pi/2$, however, the exact positions of the zeros do not 
coincide. Using a leading order Taylor expansion near $\mu_I/T=\pi/2$ we get the 
zero for $\chi^B_2$ at the position $\mu_I \approx \frac{\pi}{2} - {4 p_2(T)}/{p_1(T)}$ while the zero for $\chi^B_4$ to be at
the position $\mu_I \approx \frac{\pi}{2} - {16 p_2(T)}/{p_1(T)}$. This leads to a divergence in $\chi^B_4/\chi^B_2$ at $\mu_I \approx \frac{\pi}{2} - {4 p_2(T)}/{p_1(T)}$. This divergence is physical, but it is 
a property of a weakly interacting gas of hadrons and 
has nothing to do with criticality. Thus, while it limits the convergence of the
Taylor expansion of $\chi^B_4/\chi^B_2$, this is only an inconvenience when one
tries to look for the critical endpoint. Note, however, that this divergence in
$\chi^B_4/\chi^B_2$ at imaginary $\mu_B$ does not limit the radius of convergence
of the Taylor expansions of $\chi^B_2$ and $\chi^B_4$ themselves and thus taking the ratio of the Taylor coefficients, instead of the Taylor expansion of the ratio is more reasonable.

\section{Behavior of the denominator}

\begin{figure}[t]
    \centering
    \includegraphics[width=0.8\linewidth]{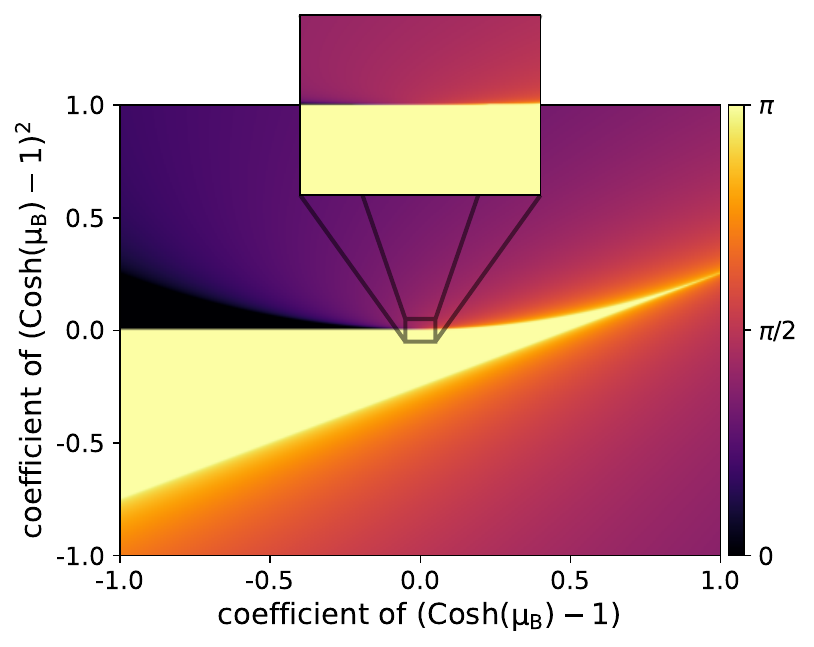}
    \caption{
    \label{fig:denominator}
    Imaginary part of the zeros of the denominator
    for the Roberge-Weiss symmetric Padé approximant used in our study as a function
    of the $q_1$ and $q_2$ coefficients. A zoomed
    in version near the origin is also shown on top.
}
\end{figure}

The denominator for the Padé approximants used in our study reads
\begin{equation*}
1+ q_1 (\cosh \hat \mu_B - 1) + q_2 (\cosh \hat \mu_B -1)^2 \rm.
\end{equation*}
The estimated Lee-Yang zeros are given by the zeros of this denominator. The imaginary part
of these zeros is then used for the extrapolation
to the critical point. Since the ansatz respects
Roberge-Weiss symmetry, the imaginary part of the
Lee-Yang zeros is restricted to be between zero
and $\pi$, after which the function periodically
repeats itself. The imaginary part as a function
of the coefficients $q_1$ and $q_2$ is shown in Fig.~\ref{fig:denominator}. On can see that
the function actually takes all values between
zero and $\pi$ and thus is in principle able
to resolve Lee-Yang zeros close to the origin.
A zoom in version near the origin is also shown on top. Here, we see the instability between the
$\pi$ and approximately $\pi/2$ values mentioned
in the main text, depending on the sign of the coefficient multiplying $(\cosh(\hat \mu_B)-1)^2$.

\section{Statistics \label{app:stat}}

\begin{table*}
\begin{tabular}{|c|ccc|rrrrrrrrr|}
\hline$T$ [MeV] & $\beta$ & $m_l$ & $m_s$ &\multicolumn{9}{|c|}{ \# configurations at $\mu_B=i\pi(j/8)$}\\
\hline
&\multicolumn{3}{|c|}{$16^3\times8$ lattice}&$j=0$ &$j=1$ &$j=2$ &$j=3$ &$j=4$ &$j=5$ &$j=6$ &$j=7$ &$j=8$ \\
\hline
100 & 0.4846 & 0.00490631 & 0.1355610 & 915010&-&-&-&-&-&-&-&-\\
105 & 0.5050 & 0.00458740 & 0.1267500 & 894220&-&-&-&-&-&-&-&-\\
110 & 0.5236 & 0.00432111 & 0.1193920 & 760795&-&-&-&-&-&-&-&-\\
115 & 0.5406 & 0.00409845 & 0.1132400 & 1054502&-&-&-&-&-&-&-&-\\
120 & 0.5560 & 0.00390982 & 0.1080280 & 1082567&-&-&-&-&-&-&-&-\\
125 & 0.5700 & 0.00374705 & 0.1035310 & 1726868&-&-&-&-&-&-&-&-\\
130 & 0.5829 & 0.00360381 & 0.0995733 & 2132329&-&96440&-&96082&-&95775&-&95625\\
135 & 0.5947 & 0.00347548 & 0.0960274 & 1453169&110859&105287&104348&120205&103744&103749&103196&103084\\
140 & 0.6056 & 0.00335869 & 0.0928007 & 4010320&99904&105232&104911&359058&103501&99483&102589&113865\\
145 & 0.6158 & 0.00325107 & 0.0898270 & 2579944&106703&106386&105027&121944&103958&103239&102471&115049\\
150 & 0.6252 & 0.00315088 & 0.0870590 & 1688860&114258&114018&112067&117290&109804&97932&96239&109087\\
155 & 0.6341 & 0.00305689 & 0.0844619 & 1623503&109132&108649&107158&125765&104061&102762&101464&115855\\
160 & 0.6425 & 0.00296817 & 0.0820105 & 2229437&112026&111086&108615&128345&104466&95570&101203&108589\\
165 & 0.6504 & 0.00288403 & 0.0796857 & 1768242&103692&101652&89855&121047&95726&92934&86765&105389\\
170 & 0.6579 & 0.00280394 & 0.0774727 & 1204210&\multicolumn{8}{c|}{-}\\
175 & 0.6651 & 0.00272748 & 0.0753604 & 1602383&\multicolumn{8}{c|}{-}\\
180 & 0.6719 & 0.00265434 & 0.0733394 & 1357330&\multicolumn{8}{c|}{-}\\
185 & 0.6785 & 0.00258424 & 0.0714026 &  855048&\multicolumn{8}{c|}{-}\\
190 & 0.6848 & 0.00251696 & 0.0695436 &  318710&\multicolumn{8}{c|}{-}\\
195 & 0.6909 & 0.00245231 & 0.0677573 &  362672&\multicolumn{8}{c|}{-}\\
200 & 0.6968 & 0.00239013 & 0.0660393 &  324768&\multicolumn{8}{c|}{-}\\
205 & 0.7025 & 0.00233028 & 0.0643856 &  159500&\multicolumn{8}{c|}{-}\\
210 & 0.7080 & 0.00227263 & 0.0627928 &  260084&\multicolumn{8}{c|}{-}\\
220 & 0.7188 & 0.00216352 & 0.0597782 &  249068&\multicolumn{8}{c|}{-}\\
230 & 0.7290 & 0.00206205 & 0.0569745 &  257480&\multicolumn{8}{c|}{-}\\
240 & 0.7389 & 0.00196759 & 0.0543644 &  270856&\multicolumn{8}{c|}{-}\\
250 & 0.7485 & 0.00187961 & 0.0519336 &  430495&\multicolumn{8}{c|}{-}\\
260 & 0.7580 & 0.00179768 & 0.0496700 &  281616&\multicolumn{8}{c|}{-}\\
280 & 0.7765 & 0.00165052 & 0.0456040 &  289961&\multicolumn{8}{c|}{-}\\
300 & 0.7946 & 0.00152351 & 0.0420946 &  297512&\multicolumn{8}{c|}{-}\\
\hline
\end{tabular}
    \caption{\label{tab:stat}
    Simulation parameters and statistics of the used ensembles. The
    configurations were separated by 20 RHMC updates. On each configuration a
    full eigenvalue analysis has been carried out at both the light and strange
    masses. }
\end{table*}


\begin{thebibliography}{52}%
\makeatletter
\providecommand \@ifxundefined [1]{%
 \@ifx{#1\undefined}
}%
\providecommand \@ifnum [1]{%
 \ifnum #1\expandafter \@firstoftwo
 \else \expandafter \@secondoftwo
 \fi
}%
\providecommand \@ifx [1]{%
 \ifx #1\expandafter \@firstoftwo
 \else \expandafter \@secondoftwo
 \fi
}%
\providecommand \natexlab [1]{#1}%
\providecommand \enquote  [1]{``#1''}%
\providecommand \bibnamefont  [1]{#1}%
\providecommand \bibfnamefont [1]{#1}%
\providecommand \citenamefont [1]{#1}%
\providecommand \href@noop [0]{\@secondoftwo}%
\providecommand \href [0]{\begingroup \@sanitize@url \@href}%
\providecommand \@href[1]{\@@startlink{#1}\@@href}%
\providecommand \@@href[1]{\endgroup#1\@@endlink}%
\providecommand \@sanitize@url [0]{\catcode `\\12\catcode `\$12\catcode `\&12\catcode `\#12\catcode `\^12\catcode `\_12\catcode `\%12\relax}%
\providecommand \@@startlink[1]{}%
\providecommand \@@endlink[0]{}%
\providecommand \url  [0]{\begingroup\@sanitize@url \@url }%
\providecommand \@url [1]{\endgroup\@href {#1}{\urlprefix }}%
\providecommand \urlprefix  [0]{URL }%
\providecommand \Eprint [0]{\href }%
\providecommand \doibase [0]{http://dx.doi.org/}%
\providecommand \selectlanguage [0]{\@gobble}%
\providecommand \bibinfo  [0]{\@secondoftwo}%
\providecommand \bibfield  [0]{\@secondoftwo}%
\providecommand \translation [1]{[#1]}%
\providecommand \BibitemOpen [0]{}%
\providecommand \bibitemStop [0]{}%
\providecommand \bibitemNoStop [0]{.\EOS\space}%
\providecommand \EOS [0]{\spacefactor3000\relax}%
\providecommand \BibitemShut  [1]{\csname bibitem#1\endcsname}%
\let\auto@bib@innerbib\@empty
\bibitem [{\citenamefont {Aoki}\ \emph {et~al.}(2006)\citenamefont {Aoki}, \citenamefont {Endrodi}, \citenamefont {Fodor}, \citenamefont {Katz},\ and\ \citenamefont {Szabo}}]{Aoki:2006we}%
  \BibitemOpen
  \bibfield  {author} {\bibinfo {author} {\bibfnamefont {Y.}~\bibnamefont {Aoki}}, \bibinfo {author} {\bibfnamefont {G.}~\bibnamefont {Endrodi}}, \bibinfo {author} {\bibfnamefont {Z.}~\bibnamefont {Fodor}}, \bibinfo {author} {\bibfnamefont {S.}~\bibnamefont {Katz}}, \ and\ \bibinfo {author} {\bibfnamefont {K.}~\bibnamefont {Szabo}},\ }\href {\doibase 10.1038/nature05120} {\bibfield  {journal} {\bibinfo  {journal} {Nature}\ }\textbf {\bibinfo {volume} {443}},\ \bibinfo {pages} {675} (\bibinfo {year} {2006})},\ \Eprint {http://arxiv.org/abs/hep-lat/0611014} {arXiv:hep-lat/0611014 [hep-lat]} \BibitemShut {NoStop}%
\bibitem [{\citenamefont {Bonati}\ \emph {et~al.}(2018)\citenamefont {Bonati}, \citenamefont {D'Elia}, \citenamefont {Negro}, \citenamefont {Sanfilippo},\ and\ \citenamefont {Zambello}}]{Bonati:2018nut}%
  \BibitemOpen
  \bibfield  {author} {\bibinfo {author} {\bibfnamefont {C.}~\bibnamefont {Bonati}}, \bibinfo {author} {\bibfnamefont {M.}~\bibnamefont {D'Elia}}, \bibinfo {author} {\bibfnamefont {F.}~\bibnamefont {Negro}}, \bibinfo {author} {\bibfnamefont {F.}~\bibnamefont {Sanfilippo}}, \ and\ \bibinfo {author} {\bibfnamefont {K.}~\bibnamefont {Zambello}},\ }\href {\doibase 10.1103/PhysRevD.98.054510} {\bibfield  {journal} {\bibinfo  {journal} {Phys. Rev.}\ }\textbf {\bibinfo {volume} {D98}},\ \bibinfo {pages} {054510} (\bibinfo {year} {2018})},\ \Eprint {http://arxiv.org/abs/1805.02960} {arXiv:1805.02960 [hep-lat]} \BibitemShut {NoStop}%
\bibitem [{\citenamefont {Bazavov}\ \emph {et~al.}(2019)\citenamefont {Bazavov} \emph {et~al.}}]{HotQCD:2018pds}%
  \BibitemOpen
  \bibfield  {author} {\bibinfo {author} {\bibfnamefont {A.}~\bibnamefont {Bazavov}} \emph {et~al.} (\bibinfo {collaboration} {HotQCD}),\ }\href {\doibase 10.1016/j.physletb.2019.05.013} {\bibfield  {journal} {\bibinfo  {journal} {Phys. Lett. B}\ }\textbf {\bibinfo {volume} {795}},\ \bibinfo {pages} {15} (\bibinfo {year} {2019})},\ \Eprint {http://arxiv.org/abs/1812.08235} {arXiv:1812.08235 [hep-lat]} \BibitemShut {NoStop}%
\bibitem [{\citenamefont {Borsanyi}\ \emph {et~al.}(2020)\citenamefont {Borsanyi}, \citenamefont {Fodor}, \citenamefont {Guenther}, \citenamefont {Kara}, \citenamefont {Katz}, \citenamefont {Parotto}, \citenamefont {Pasztor}, \citenamefont {Ratti},\ and\ \citenamefont {Szabo}}]{Borsanyi:2020fev}%
  \BibitemOpen
  \bibfield  {author} {\bibinfo {author} {\bibfnamefont {S.}~\bibnamefont {Borsanyi}}, \bibinfo {author} {\bibfnamefont {Z.}~\bibnamefont {Fodor}}, \bibinfo {author} {\bibfnamefont {J.~N.}\ \bibnamefont {Guenther}}, \bibinfo {author} {\bibfnamefont {R.}~\bibnamefont {Kara}}, \bibinfo {author} {\bibfnamefont {S.~D.}\ \bibnamefont {Katz}}, \bibinfo {author} {\bibfnamefont {P.}~\bibnamefont {Parotto}}, \bibinfo {author} {\bibfnamefont {A.}~\bibnamefont {Pasztor}}, \bibinfo {author} {\bibfnamefont {C.}~\bibnamefont {Ratti}}, \ and\ \bibinfo {author} {\bibfnamefont {K.~K.}\ \bibnamefont {Szabo}},\ }\href@noop {} {\bibfield  {journal} {\bibinfo  {journal} {Phys. Rev. Lett.}\ }\textbf {\bibinfo {volume} {125}},\ \bibinfo {pages} {052001} (\bibinfo {year} {2020})},\ \Eprint {http://arxiv.org/abs/2002.02821} {arXiv:2002.02821 [hep-lat]} \BibitemShut {NoStop}%
\bibitem [{\citenamefont {Bollweg}\ \emph {et~al.}(2022)\citenamefont {Bollweg}, \citenamefont {Goswami}, \citenamefont {Kaczmarek}, \citenamefont {Karsch}, \citenamefont {Mukherjee}, \citenamefont {Petreczky}, \citenamefont {Schmidt},\ and\ \citenamefont {Scior}}]{Bollweg:2022rps}%
  \BibitemOpen
  \bibfield  {author} {\bibinfo {author} {\bibfnamefont {D.}~\bibnamefont {Bollweg}}, \bibinfo {author} {\bibfnamefont {J.}~\bibnamefont {Goswami}}, \bibinfo {author} {\bibfnamefont {O.}~\bibnamefont {Kaczmarek}}, \bibinfo {author} {\bibfnamefont {F.}~\bibnamefont {Karsch}}, \bibinfo {author} {\bibfnamefont {S.}~\bibnamefont {Mukherjee}}, \bibinfo {author} {\bibfnamefont {P.}~\bibnamefont {Petreczky}}, \bibinfo {author} {\bibfnamefont {C.}~\bibnamefont {Schmidt}}, \ and\ \bibinfo {author} {\bibfnamefont {P.}~\bibnamefont {Scior}} (\bibinfo {collaboration} {HotQCD}),\ }\href {\doibase 10.1103/PhysRevD.105.074511} {\bibfield  {journal} {\bibinfo  {journal} {Phys. Rev. D}\ }\textbf {\bibinfo {volume} {105}},\ \bibinfo {pages} {074511} (\bibinfo {year} {2022})},\ \Eprint {http://arxiv.org/abs/2202.09184} {arXiv:2202.09184 [hep-lat]} \BibitemShut {NoStop}%
\bibitem [{\citenamefont {Borsanyi}\ \emph {et~al.}(2024{\natexlab{a}})\citenamefont {Borsanyi}, \citenamefont {Fodor}, \citenamefont {Guenther}, \citenamefont {Katz}, \citenamefont {Parotto}, \citenamefont {Pasztor}, \citenamefont {Pesznyak}, \citenamefont {Szabo},\ and\ \citenamefont {Wong}}]{Borsanyi:2023wno}%
  \BibitemOpen
  \bibfield  {author} {\bibinfo {author} {\bibfnamefont {S.}~\bibnamefont {Borsanyi}}, \bibinfo {author} {\bibfnamefont {Z.}~\bibnamefont {Fodor}}, \bibinfo {author} {\bibfnamefont {J.~N.}\ \bibnamefont {Guenther}}, \bibinfo {author} {\bibfnamefont {S.~D.}\ \bibnamefont {Katz}}, \bibinfo {author} {\bibfnamefont {P.}~\bibnamefont {Parotto}}, \bibinfo {author} {\bibfnamefont {A.}~\bibnamefont {Pasztor}}, \bibinfo {author} {\bibfnamefont {D.}~\bibnamefont {Pesznyak}}, \bibinfo {author} {\bibfnamefont {K.~K.}\ \bibnamefont {Szabo}}, \ and\ \bibinfo {author} {\bibfnamefont {C.~H.}\ \bibnamefont {Wong}},\ }\href {\doibase 10.1103/PhysRevD.110.L011501} {\bibfield  {journal} {\bibinfo  {journal} {Phys. Rev. D}\ }\textbf {\bibinfo {volume} {110}},\ \bibinfo {pages} {L011501} (\bibinfo {year} {2024}{\natexlab{a}})},\ \Eprint {http://arxiv.org/abs/2312.07528} {arXiv:2312.07528 [hep-lat]} \BibitemShut {NoStop}%
\bibitem [{\citenamefont {Borsanyi}\ \emph {et~al.}(2018)\citenamefont {Borsanyi}, \citenamefont {Fodor}, \citenamefont {Guenther}, \citenamefont {Katz}, \citenamefont {Szabo}, \citenamefont {Pasztor}, \citenamefont {Portillo},\ and\ \citenamefont {Ratti}}]{Borsanyi:2018grb}%
  \BibitemOpen
  \bibfield  {author} {\bibinfo {author} {\bibfnamefont {S.}~\bibnamefont {Borsanyi}}, \bibinfo {author} {\bibfnamefont {Z.}~\bibnamefont {Fodor}}, \bibinfo {author} {\bibfnamefont {J.~N.}\ \bibnamefont {Guenther}}, \bibinfo {author} {\bibfnamefont {S.~K.}\ \bibnamefont {Katz}}, \bibinfo {author} {\bibfnamefont {K.~K.}\ \bibnamefont {Szabo}}, \bibinfo {author} {\bibfnamefont {A.}~\bibnamefont {Pasztor}}, \bibinfo {author} {\bibfnamefont {I.}~\bibnamefont {Portillo}}, \ and\ \bibinfo {author} {\bibfnamefont {C.}~\bibnamefont {Ratti}},\ }\href {\doibase 10.1007/JHEP10(2018)205} {\bibfield  {journal} {\bibinfo  {journal} {JHEP}\ }\textbf {\bibinfo {volume} {10}},\ \bibinfo {pages} {205} (\bibinfo {year} {2018})},\ \Eprint {http://arxiv.org/abs/1805.04445} {arXiv:1805.04445 [hep-lat]} \BibitemShut {NoStop}%
\bibitem [{\citenamefont {Kov\'acs}\ \emph {et~al.}(2016)\citenamefont {Kov\'acs}, \citenamefont {Sz\'ep},\ and\ \citenamefont {Wolf}}]{Kovacs:2016juc}%
  \BibitemOpen
  \bibfield  {author} {\bibinfo {author} {\bibfnamefont {P.}~\bibnamefont {Kov\'acs}}, \bibinfo {author} {\bibfnamefont {Z.}~\bibnamefont {Sz\'ep}}, \ and\ \bibinfo {author} {\bibfnamefont {G.}~\bibnamefont {Wolf}},\ }\href {\doibase 10.1103/PhysRevD.93.114014} {\bibfield  {journal} {\bibinfo  {journal} {Phys. Rev. D}\ }\textbf {\bibinfo {volume} {93}},\ \bibinfo {pages} {114014} (\bibinfo {year} {2016})},\ \Eprint {http://arxiv.org/abs/1601.05291} {arXiv:1601.05291 [hep-ph]} \BibitemShut {NoStop}%
\bibitem [{\citenamefont {Fu}\ \emph {et~al.}(2025)\citenamefont {Fu}, \citenamefont {Luo}, \citenamefont {Pawlowski}, \citenamefont {Rennecke},\ and\ \citenamefont {Yin}}]{Fu:2023lcm}%
  \BibitemOpen
  \bibfield  {author} {\bibinfo {author} {\bibfnamefont {W.-j.}\ \bibnamefont {Fu}}, \bibinfo {author} {\bibfnamefont {X.}~\bibnamefont {Luo}}, \bibinfo {author} {\bibfnamefont {J.~M.}\ \bibnamefont {Pawlowski}}, \bibinfo {author} {\bibfnamefont {F.}~\bibnamefont {Rennecke}}, \ and\ \bibinfo {author} {\bibfnamefont {S.}~\bibnamefont {Yin}},\ }\href {\doibase 10.1103/PhysRevD.111.L031502} {\bibfield  {journal} {\bibinfo  {journal} {Phys. Rev. D}\ }\textbf {\bibinfo {volume} {111}},\ \bibinfo {pages} {L031502} (\bibinfo {year} {2025})},\ \Eprint {http://arxiv.org/abs/2308.15508} {arXiv:2308.15508 [hep-ph]} \BibitemShut {NoStop}%
\bibitem [{\citenamefont {Isserstedt}\ \emph {et~al.}(2019)\citenamefont {Isserstedt}, \citenamefont {Buballa}, \citenamefont {Fischer},\ and\ \citenamefont {Gunkel}}]{Isserstedt:2019pgx}%
  \BibitemOpen
  \bibfield  {author} {\bibinfo {author} {\bibfnamefont {P.}~\bibnamefont {Isserstedt}}, \bibinfo {author} {\bibfnamefont {M.}~\bibnamefont {Buballa}}, \bibinfo {author} {\bibfnamefont {C.~S.}\ \bibnamefont {Fischer}}, \ and\ \bibinfo {author} {\bibfnamefont {P.~J.}\ \bibnamefont {Gunkel}},\ }\href {\doibase 10.1103/PhysRevD.100.074011} {\bibfield  {journal} {\bibinfo  {journal} {Phys. Rev.}\ }\textbf {\bibinfo {volume} {D100}},\ \bibinfo {pages} {074011} (\bibinfo {year} {2019})},\ \Eprint {http://arxiv.org/abs/1906.11644} {arXiv:1906.11644 [hep-ph]} \BibitemShut {NoStop}%
\bibitem [{\citenamefont {Gao}\ and\ \citenamefont {Pawlowski}(2020)}]{Gao:2020qsj}%
  \BibitemOpen
  \bibfield  {author} {\bibinfo {author} {\bibfnamefont {F.}~\bibnamefont {Gao}}\ and\ \bibinfo {author} {\bibfnamefont {J.~M.}\ \bibnamefont {Pawlowski}},\ }\href {\doibase 10.1103/PhysRevD.102.034027} {\bibfield  {journal} {\bibinfo  {journal} {Phys. Rev. D}\ }\textbf {\bibinfo {volume} {102}},\ \bibinfo {pages} {034027} (\bibinfo {year} {2020})},\ \Eprint {http://arxiv.org/abs/2002.07500} {arXiv:2002.07500 [hep-ph]} \BibitemShut {NoStop}%
\bibitem [{\citenamefont {Hippert}\ \emph {et~al.}(2024)\citenamefont {Hippert}, \citenamefont {Grefa}, \citenamefont {Manning}, \citenamefont {Noronha}, \citenamefont {Noronha-Hostler}, \citenamefont {Portillo~Vazquez}, \citenamefont {Ratti}, \citenamefont {Rougemont},\ and\ \citenamefont {Trujillo}}]{Hippert:2023bel}%
  \BibitemOpen
  \bibfield  {author} {\bibinfo {author} {\bibfnamefont {M.}~\bibnamefont {Hippert}}, \bibinfo {author} {\bibfnamefont {J.}~\bibnamefont {Grefa}}, \bibinfo {author} {\bibfnamefont {T.~A.}\ \bibnamefont {Manning}}, \bibinfo {author} {\bibfnamefont {J.}~\bibnamefont {Noronha}}, \bibinfo {author} {\bibfnamefont {J.}~\bibnamefont {Noronha-Hostler}}, \bibinfo {author} {\bibfnamefont {I.}~\bibnamefont {Portillo~Vazquez}}, \bibinfo {author} {\bibfnamefont {C.}~\bibnamefont {Ratti}}, \bibinfo {author} {\bibfnamefont {R.}~\bibnamefont {Rougemont}}, \ and\ \bibinfo {author} {\bibfnamefont {M.}~\bibnamefont {Trujillo}},\ }\href {\doibase 10.1103/PhysRevD.110.094006} {\bibfield  {journal} {\bibinfo  {journal} {Phys. Rev. D}\ }\textbf {\bibinfo {volume} {110}},\ \bibinfo {pages} {094006} (\bibinfo {year} {2024})},\ \Eprint {http://arxiv.org/abs/2309.00579} {arXiv:2309.00579 [nucl-th]} \BibitemShut {NoStop}%
\bibitem [{\citenamefont {Bors\'anyi}\ \emph {et~al.}(2021)\citenamefont {Bors\'anyi}, \citenamefont {Fodor}, \citenamefont {Guenther}, \citenamefont {Kara}, \citenamefont {Katz}, \citenamefont {Parotto}, \citenamefont {P\'asztor}, \citenamefont {Ratti},\ and\ \citenamefont {Szab\'o}}]{Borsanyi:2021sxv}%
  \BibitemOpen
  \bibfield  {author} {\bibinfo {author} {\bibfnamefont {S.}~\bibnamefont {Bors\'anyi}}, \bibinfo {author} {\bibfnamefont {Z.}~\bibnamefont {Fodor}}, \bibinfo {author} {\bibfnamefont {J.~N.}\ \bibnamefont {Guenther}}, \bibinfo {author} {\bibfnamefont {R.}~\bibnamefont {Kara}}, \bibinfo {author} {\bibfnamefont {S.~D.}\ \bibnamefont {Katz}}, \bibinfo {author} {\bibfnamefont {P.}~\bibnamefont {Parotto}}, \bibinfo {author} {\bibfnamefont {A.}~\bibnamefont {P\'asztor}}, \bibinfo {author} {\bibfnamefont {C.}~\bibnamefont {Ratti}}, \ and\ \bibinfo {author} {\bibfnamefont {K.~K.}\ \bibnamefont {Szab\'o}},\ }\href {\doibase 10.1103/PhysRevLett.126.232001} {\bibfield  {journal} {\bibinfo  {journal} {Phys. Rev. Lett.}\ }\textbf {\bibinfo {volume} {126}},\ \bibinfo {pages} {232001} (\bibinfo {year} {2021})},\ \Eprint {http://arxiv.org/abs/2102.06660} {arXiv:2102.06660 [hep-lat]} \BibitemShut {NoStop}%
\bibitem [{\citenamefont {Borsanyi}\ \emph {et~al.}(2022)\citenamefont {Borsanyi}, \citenamefont {Fodor}, \citenamefont {Guenther}, \citenamefont {Kara}, \citenamefont {Parotto}, \citenamefont {Pasztor}, \citenamefont {Ratti},\ and\ \citenamefont {Szabo}}]{Borsanyi:2022qlh}%
  \BibitemOpen
  \bibfield  {author} {\bibinfo {author} {\bibfnamefont {S.}~\bibnamefont {Borsanyi}}, \bibinfo {author} {\bibfnamefont {Z.}~\bibnamefont {Fodor}}, \bibinfo {author} {\bibfnamefont {J.~N.}\ \bibnamefont {Guenther}}, \bibinfo {author} {\bibfnamefont {R.}~\bibnamefont {Kara}}, \bibinfo {author} {\bibfnamefont {P.}~\bibnamefont {Parotto}}, \bibinfo {author} {\bibfnamefont {A.}~\bibnamefont {Pasztor}}, \bibinfo {author} {\bibfnamefont {C.}~\bibnamefont {Ratti}}, \ and\ \bibinfo {author} {\bibfnamefont {K.~K.}\ \bibnamefont {Szabo}},\ }\href {\doibase 10.1103/PhysRevD.105.114504} {\bibfield  {journal} {\bibinfo  {journal} {Phys. Rev. D}\ }\textbf {\bibinfo {volume} {105}},\ \bibinfo {pages} {114504} (\bibinfo {year} {2022})},\ \Eprint {http://arxiv.org/abs/2202.05574} {arXiv:2202.05574 [hep-lat]} \BibitemShut {NoStop}%
\bibitem [{\citenamefont {Kahangirwe}\ \emph {et~al.}(2024)\citenamefont {Kahangirwe}, \citenamefont {Gonzalez}, \citenamefont {Mu\~noz}, \citenamefont {Ratti},\ and\ \citenamefont {Vovchenko}}]{Kahangirwe:2024xyl}%
  \BibitemOpen
  \bibfield  {author} {\bibinfo {author} {\bibfnamefont {M.}~\bibnamefont {Kahangirwe}}, \bibinfo {author} {\bibfnamefont {I.}~\bibnamefont {Gonzalez}}, \bibinfo {author} {\bibfnamefont {J.~A.}\ \bibnamefont {Mu\~noz}}, \bibinfo {author} {\bibfnamefont {C.}~\bibnamefont {Ratti}}, \ and\ \bibinfo {author} {\bibfnamefont {V.}~\bibnamefont {Vovchenko}},\ }\href@noop {} {\  (\bibinfo {year} {2024})},\ \Eprint {http://arxiv.org/abs/2408.04588} {arXiv:2408.04588 [nucl-th]} \BibitemShut {NoStop}%
\bibitem [{\citenamefont {Wen}\ \emph {et~al.}(2024)\citenamefont {Wen}, \citenamefont {Yin},\ and\ \citenamefont {Fu}}]{Wen:2024hbz}%
  \BibitemOpen
  \bibfield  {author} {\bibinfo {author} {\bibfnamefont {R.}~\bibnamefont {Wen}}, \bibinfo {author} {\bibfnamefont {S.}~\bibnamefont {Yin}}, \ and\ \bibinfo {author} {\bibfnamefont {W.-j.}\ \bibnamefont {Fu}},\ }\href {\doibase 10.1103/PhysRevD.110.016008} {\bibfield  {journal} {\bibinfo  {journal} {Phys. Rev. D}\ }\textbf {\bibinfo {volume} {110}},\ \bibinfo {pages} {016008} (\bibinfo {year} {2024})},\ \Eprint {http://arxiv.org/abs/2403.06770} {arXiv:2403.06770 [hep-ph]} \BibitemShut {NoStop}%
\bibitem [{\citenamefont {Abuali}\ \emph {et~al.}(2025)\citenamefont {Abuali}, \citenamefont {Bors{\'a}nyi}, \citenamefont {Fodor}, \citenamefont {Jahan}, \citenamefont {Kahangirwe}, \citenamefont {Parotto}, \citenamefont {P{\'a}sztor}, \citenamefont {Ratti}, \citenamefont {Shah},\ and\ \citenamefont {Trabulsi}}]{Abuali:2025tbd}%
  \BibitemOpen
  \bibfield  {author} {\bibinfo {author} {\bibfnamefont {A.}~\bibnamefont {Abuali}}, \bibinfo {author} {\bibfnamefont {S.}~\bibnamefont {Bors{\'a}nyi}}, \bibinfo {author} {\bibfnamefont {Z.}~\bibnamefont {Fodor}}, \bibinfo {author} {\bibfnamefont {J.}~\bibnamefont {Jahan}}, \bibinfo {author} {\bibfnamefont {M.}~\bibnamefont {Kahangirwe}}, \bibinfo {author} {\bibfnamefont {P.}~\bibnamefont {Parotto}}, \bibinfo {author} {\bibfnamefont {A.}~\bibnamefont {P{\'a}sztor}}, \bibinfo {author} {\bibfnamefont {C.}~\bibnamefont {Ratti}}, \bibinfo {author} {\bibfnamefont {H.}~\bibnamefont {Shah}}, \ and\ \bibinfo {author} {\bibfnamefont {S.~A.}\ \bibnamefont {Trabulsi}},\ }\href@noop {} {\  (\bibinfo {year} {2025})},\ \Eprint {http://arxiv.org/abs/2504.01881} {arXiv:2504.01881 [hep-lat]} \BibitemShut {NoStop}%
\bibitem [{\citenamefont {Shah}\ \emph {et~al.}(2024)\citenamefont {Shah}, \citenamefont {Hippert}, \citenamefont {Noronha}, \citenamefont {Ratti},\ and\ \citenamefont {Vovchenko}}]{Shah:2024img}%
  \BibitemOpen
  \bibfield  {author} {\bibinfo {author} {\bibfnamefont {H.}~\bibnamefont {Shah}}, \bibinfo {author} {\bibfnamefont {M.}~\bibnamefont {Hippert}}, \bibinfo {author} {\bibfnamefont {J.}~\bibnamefont {Noronha}}, \bibinfo {author} {\bibfnamefont {C.}~\bibnamefont {Ratti}}, \ and\ \bibinfo {author} {\bibfnamefont {V.}~\bibnamefont {Vovchenko}},\ }\href@noop {} {\  (\bibinfo {year} {2024})},\ \Eprint {http://arxiv.org/abs/2410.16206} {arXiv:2410.16206 [hep-ph]} \BibitemShut {NoStop}%
\bibitem [{\citenamefont {Borsanyi}\ \emph {et~al.}(2025)\citenamefont {Borsanyi}, \citenamefont {Fodor}, \citenamefont {Guenther}, \citenamefont {Parotto}, \citenamefont {Pasztor}, \citenamefont {Ratti}, \citenamefont {Vovchenko},\ and\ \citenamefont {Wong}}]{Borsanyi:2025dyp}%
  \BibitemOpen
  \bibfield  {author} {\bibinfo {author} {\bibfnamefont {S.}~\bibnamefont {Borsanyi}}, \bibinfo {author} {\bibfnamefont {Z.}~\bibnamefont {Fodor}}, \bibinfo {author} {\bibfnamefont {J.~N.}\ \bibnamefont {Guenther}}, \bibinfo {author} {\bibfnamefont {P.}~\bibnamefont {Parotto}}, \bibinfo {author} {\bibfnamefont {A.}~\bibnamefont {Pasztor}}, \bibinfo {author} {\bibfnamefont {C.}~\bibnamefont {Ratti}}, \bibinfo {author} {\bibfnamefont {V.}~\bibnamefont {Vovchenko}}, \ and\ \bibinfo {author} {\bibfnamefont {C.~H.}\ \bibnamefont {Wong}},\ }\href@noop {} {\  (\bibinfo {year} {2025})},\ \Eprint {http://arxiv.org/abs/2502.10267} {arXiv:2502.10267 [hep-lat]} \BibitemShut {NoStop}%
\bibitem [{\citenamefont {Marczenko}\ \emph {et~al.}(2025)\citenamefont {Marczenko}, \citenamefont {Szyma{\'n}ski},\ and\ \citenamefont {Kov{\'a}cs}}]{Marczenko:2025znt}%
  \BibitemOpen
  \bibfield  {author} {\bibinfo {author} {\bibfnamefont {M.}~\bibnamefont {Marczenko}}, \bibinfo {author} {\bibfnamefont {M.}~\bibnamefont {Szyma{\'n}ski}}, \ and\ \bibinfo {author} {\bibfnamefont {G.}~\bibnamefont {Kov{\'a}cs}},\ }\href@noop {} {\  (\bibinfo {year} {2025})},\ \Eprint {http://arxiv.org/abs/2504.04446} {arXiv:2504.04446 [hep-ph]} \BibitemShut {NoStop}%
\bibitem [{\citenamefont {Fodor}\ and\ \citenamefont {Katz}(2004)}]{Fodor:2004nz}%
  \BibitemOpen
  \bibfield  {author} {\bibinfo {author} {\bibfnamefont {Z.}~\bibnamefont {Fodor}}\ and\ \bibinfo {author} {\bibfnamefont {S.}~\bibnamefont {Katz}},\ }\href {\doibase 10.1088/1126-6708/2004/04/050} {\bibfield  {journal} {\bibinfo  {journal} {JHEP}\ }\textbf {\bibinfo {volume} {0404}},\ \bibinfo {pages} {050} (\bibinfo {year} {2004})},\ \Eprint {http://arxiv.org/abs/hep-lat/0402006} {arXiv:hep-lat/0402006 [hep-lat]} \BibitemShut {NoStop}%
\bibitem [{\citenamefont {Giordano}\ and\ \citenamefont {Psztor}(2019)}]{Giordano:2019slo}%
  \BibitemOpen
  \bibfield  {author} {\bibinfo {author} {\bibfnamefont {M.}~\bibnamefont {Giordano}}\ and\ \bibinfo {author} {\bibfnamefont {A.}~\bibnamefont {Psztor}},\ }\href {\doibase 10.1103/PhysRevD.99.114510} {\bibfield  {journal} {\bibinfo  {journal} {Phys. Rev.}\ }\textbf {\bibinfo {volume} {D99}},\ \bibinfo {pages} {114510} (\bibinfo {year} {2019})},\ \Eprint {http://arxiv.org/abs/1904.01974} {arXiv:1904.01974 [hep-lat]} \BibitemShut {NoStop}%
\bibitem [{\citenamefont {Giordano}\ \emph {et~al.}(2020)\citenamefont {Giordano}, \citenamefont {Kapas}, \citenamefont {Katz}, \citenamefont {Nogradi},\ and\ \citenamefont {Pasztor}}]{Giordano:2019gev}%
  \BibitemOpen
  \bibfield  {author} {\bibinfo {author} {\bibfnamefont {M.}~\bibnamefont {Giordano}}, \bibinfo {author} {\bibfnamefont {K.}~\bibnamefont {Kapas}}, \bibinfo {author} {\bibfnamefont {S.~D.}\ \bibnamefont {Katz}}, \bibinfo {author} {\bibfnamefont {D.}~\bibnamefont {Nogradi}}, \ and\ \bibinfo {author} {\bibfnamefont {A.}~\bibnamefont {Pasztor}},\ }\href {\doibase 10.1103/PhysRevD.101.074511} {\bibfield  {journal} {\bibinfo  {journal} {Phys. Rev. D}\ }\textbf {\bibinfo {volume} {101}},\ \bibinfo {pages} {074511} (\bibinfo {year} {2020})},\ \Eprint {http://arxiv.org/abs/1911.00043} {arXiv:1911.00043 [hep-lat]} \BibitemShut {NoStop}%
\bibitem [{\citenamefont {Dimopoulos}\ \emph {et~al.}(2022)\citenamefont {Dimopoulos}, \citenamefont {Dini}, \citenamefont {Di~Renzo}, \citenamefont {Goswami}, \citenamefont {Nicotra}, \citenamefont {Schmidt}, \citenamefont {Singh}, \citenamefont {Zambello},\ and\ \citenamefont {Ziesch\'e}}]{Dimopoulos:2021vrk}%
  \BibitemOpen
  \bibfield  {author} {\bibinfo {author} {\bibfnamefont {P.}~\bibnamefont {Dimopoulos}}, \bibinfo {author} {\bibfnamefont {L.}~\bibnamefont {Dini}}, \bibinfo {author} {\bibfnamefont {F.}~\bibnamefont {Di~Renzo}}, \bibinfo {author} {\bibfnamefont {J.}~\bibnamefont {Goswami}}, \bibinfo {author} {\bibfnamefont {G.}~\bibnamefont {Nicotra}}, \bibinfo {author} {\bibfnamefont {C.}~\bibnamefont {Schmidt}}, \bibinfo {author} {\bibfnamefont {S.}~\bibnamefont {Singh}}, \bibinfo {author} {\bibfnamefont {K.}~\bibnamefont {Zambello}}, \ and\ \bibinfo {author} {\bibfnamefont {F.}~\bibnamefont {Ziesch\'e}},\ }\href {\doibase 10.1103/PhysRevD.105.034513} {\bibfield  {journal} {\bibinfo  {journal} {Phys. Rev. D}\ }\textbf {\bibinfo {volume} {105}},\ \bibinfo {pages} {034513} (\bibinfo {year} {2022})},\ \Eprint {http://arxiv.org/abs/2110.15933} {arXiv:2110.15933 [hep-lat]} \BibitemShut {NoStop}%
\bibitem [{\citenamefont {Basar}(2024)}]{Basar:2023nkp}%
  \BibitemOpen
  \bibfield  {author} {\bibinfo {author} {\bibfnamefont {G.}~\bibnamefont {Basar}},\ }\href {\doibase 10.1103/PhysRevC.110.015203} {\bibfield  {journal} {\bibinfo  {journal} {Phys. Rev. C}\ }\textbf {\bibinfo {volume} {110}},\ \bibinfo {pages} {015203} (\bibinfo {year} {2024})},\ \Eprint {http://arxiv.org/abs/2312.06952} {arXiv:2312.06952 [hep-th]} \BibitemShut {NoStop}%
\bibitem [{\citenamefont {Clarke}\ \emph {et~al.}(2024)\citenamefont {Clarke}, \citenamefont {Dimopoulos}, \citenamefont {Di~Renzo}, \citenamefont {Goswami}, \citenamefont {Schmidt}, \citenamefont {Singh},\ and\ \citenamefont {Zambello}}]{Clarke:2024ugt}%
  \BibitemOpen
  \bibfield  {author} {\bibinfo {author} {\bibfnamefont {D.~A.}\ \bibnamefont {Clarke}}, \bibinfo {author} {\bibfnamefont {P.}~\bibnamefont {Dimopoulos}}, \bibinfo {author} {\bibfnamefont {F.}~\bibnamefont {Di~Renzo}}, \bibinfo {author} {\bibfnamefont {J.}~\bibnamefont {Goswami}}, \bibinfo {author} {\bibfnamefont {C.}~\bibnamefont {Schmidt}}, \bibinfo {author} {\bibfnamefont {S.}~\bibnamefont {Singh}}, \ and\ \bibinfo {author} {\bibfnamefont {K.}~\bibnamefont {Zambello}},\ }\href@noop {} {\  (\bibinfo {year} {2024})},\ \Eprint {http://arxiv.org/abs/2405.10196} {arXiv:2405.10196 [hep-lat]} \BibitemShut {NoStop}%
\bibitem [{\citenamefont {Skokov}(2025)}]{Skokov:2024fac}%
  \BibitemOpen
  \bibfield  {author} {\bibinfo {author} {\bibfnamefont {V.~V.}\ \bibnamefont {Skokov}},\ }\href {\doibase 10.21468/SciPostPhysLectNotes.91} {\bibfield  {journal} {\bibinfo  {journal} {SciPost Phys. Lect. Notes}\ }\textbf {\bibinfo {volume} {91}},\ \bibinfo {pages} {1} (\bibinfo {year} {2025})},\ \Eprint {http://arxiv.org/abs/2411.02663} {arXiv:2411.02663 [hep-ph]} \BibitemShut {NoStop}%
\bibitem [{\citenamefont {Bors\'anyi}\ \emph {et~al.}(2025)\citenamefont {Bors\'anyi}, \citenamefont {Fodor}, \citenamefont {Guenther}, \citenamefont {Kara}, \citenamefont {Parotto}, \citenamefont {P\'asztor}, \citenamefont {Pirelli},\ and\ \citenamefont {Wong}}]{Borsanyi:2025lim}%
  \BibitemOpen
  \bibfield  {author} {\bibinfo {author} {\bibfnamefont {S.}~\bibnamefont {Bors\'anyi}}, \bibinfo {author} {\bibfnamefont {Z.}~\bibnamefont {Fodor}}, \bibinfo {author} {\bibfnamefont {J.~N.}\ \bibnamefont {Guenther}}, \bibinfo {author} {\bibfnamefont {R.}~\bibnamefont {Kara}}, \bibinfo {author} {\bibfnamefont {P.}~\bibnamefont {Parotto}}, \bibinfo {author} {\bibfnamefont {A.}~\bibnamefont {P\'asztor}}, \bibinfo {author} {\bibfnamefont {L.}~\bibnamefont {Pirelli}}, \ and\ \bibinfo {author} {\bibfnamefont {C.~H.}\ \bibnamefont {Wong}},\ }\href {\doibase 10.1103/PhysRevD.111.014506} {\bibfield  {journal} {\bibinfo  {journal} {Phys. Rev. D}\ }\textbf {\bibinfo {volume} {111}},\ \bibinfo {pages} {014506} (\bibinfo {year} {2025})}\BibitemShut {NoStop}%
\bibitem [{\citenamefont {Fodor}\ and\ \citenamefont {Katz}(2002{\natexlab{a}})}]{Fodor:2001au}%
  \BibitemOpen
  \bibfield  {author} {\bibinfo {author} {\bibfnamefont {Z.}~\bibnamefont {Fodor}}\ and\ \bibinfo {author} {\bibfnamefont {S.}~\bibnamefont {Katz}},\ }\href {\doibase 10.1016/S0370-2693(02)01583-6} {\bibfield  {journal} {\bibinfo  {journal} {Phys.Lett.}\ }\textbf {\bibinfo {volume} {B534}},\ \bibinfo {pages} {87} (\bibinfo {year} {2002}{\natexlab{a}})},\ \Eprint {http://arxiv.org/abs/hep-lat/0104001} {arXiv:hep-lat/0104001 [hep-lat]} \BibitemShut {NoStop}%
\bibitem [{\citenamefont {Fodor}\ and\ \citenamefont {Katz}(2002{\natexlab{b}})}]{Fodor:2001pe}%
  \BibitemOpen
  \bibfield  {author} {\bibinfo {author} {\bibfnamefont {Z.}~\bibnamefont {Fodor}}\ and\ \bibinfo {author} {\bibfnamefont {S.}~\bibnamefont {Katz}},\ }\href {\doibase 10.1088/1126-6708/2002/03/014} {\bibfield  {journal} {\bibinfo  {journal} {JHEP}\ }\textbf {\bibinfo {volume} {0203}},\ \bibinfo {pages} {014} (\bibinfo {year} {2002}{\natexlab{b}})},\ \Eprint {http://arxiv.org/abs/hep-lat/0106002} {arXiv:hep-lat/0106002 [hep-lat]} \BibitemShut {NoStop}%
\bibitem [{\citenamefont {Borsanyi}\ \emph {et~al.}(2023)\citenamefont {Borsanyi}, \citenamefont {Fodor}, \citenamefont {Giordano}, \citenamefont {Guenther}, \citenamefont {Katz}, \citenamefont {Pasztor},\ and\ \citenamefont {Wong}}]{Borsanyi:2022soo}%
  \BibitemOpen
  \bibfield  {author} {\bibinfo {author} {\bibfnamefont {S.}~\bibnamefont {Borsanyi}}, \bibinfo {author} {\bibfnamefont {Z.}~\bibnamefont {Fodor}}, \bibinfo {author} {\bibfnamefont {M.}~\bibnamefont {Giordano}}, \bibinfo {author} {\bibfnamefont {J.~N.}\ \bibnamefont {Guenther}}, \bibinfo {author} {\bibfnamefont {S.~D.}\ \bibnamefont {Katz}}, \bibinfo {author} {\bibfnamefont {A.}~\bibnamefont {Pasztor}}, \ and\ \bibinfo {author} {\bibfnamefont {C.~H.}\ \bibnamefont {Wong}},\ }\href {\doibase 10.1103/PhysRevD.107.L091503} {\bibfield  {journal} {\bibinfo  {journal} {Phys. Rev. D}\ }\textbf {\bibinfo {volume} {107}},\ \bibinfo {pages} {L091503} (\bibinfo {year} {2023})},\ \Eprint {http://arxiv.org/abs/2208.05398} {arXiv:2208.05398 [hep-lat]} \BibitemShut {NoStop}%
\bibitem [{\citenamefont {Borsanyi}\ \emph {et~al.}(2024{\natexlab{b}})\citenamefont {Borsanyi}, \citenamefont {Fodor}, \citenamefont {Giordano}, \citenamefont {Guenther}, \citenamefont {Katz}, \citenamefont {Pasztor},\ and\ \citenamefont {Wong}}]{Borsanyi:2023tdp}%
  \BibitemOpen
  \bibfield  {author} {\bibinfo {author} {\bibfnamefont {S.}~\bibnamefont {Borsanyi}}, \bibinfo {author} {\bibfnamefont {Z.}~\bibnamefont {Fodor}}, \bibinfo {author} {\bibfnamefont {M.}~\bibnamefont {Giordano}}, \bibinfo {author} {\bibfnamefont {J.~N.}\ \bibnamefont {Guenther}}, \bibinfo {author} {\bibfnamefont {S.~D.}\ \bibnamefont {Katz}}, \bibinfo {author} {\bibfnamefont {A.}~\bibnamefont {Pasztor}}, \ and\ \bibinfo {author} {\bibfnamefont {C.~H.}\ \bibnamefont {Wong}},\ }\href {\doibase 10.1103/PhysRevD.109.054509} {\bibfield  {journal} {\bibinfo  {journal} {Phys. Rev. D}\ }\textbf {\bibinfo {volume} {109}},\ \bibinfo {pages} {054509} (\bibinfo {year} {2024}{\natexlab{b}})},\ \Eprint {http://arxiv.org/abs/2308.06105} {arXiv:2308.06105 [hep-lat]} \BibitemShut {NoStop}%
\bibitem [{\citenamefont {Borsanyi}\ \emph {et~al.}(2024{\natexlab{c}})\citenamefont {Borsanyi}, \citenamefont {Fodor}, \citenamefont {Guenther}, \citenamefont {Parotto}, \citenamefont {Pasztor}, \citenamefont {Pirelli}, \citenamefont {Szabo},\ and\ \citenamefont {Wong}}]{Borsanyi:2024xrx}%
  \BibitemOpen
  \bibfield  {author} {\bibinfo {author} {\bibfnamefont {S.}~\bibnamefont {Borsanyi}}, \bibinfo {author} {\bibfnamefont {Z.}~\bibnamefont {Fodor}}, \bibinfo {author} {\bibfnamefont {J.~N.}\ \bibnamefont {Guenther}}, \bibinfo {author} {\bibfnamefont {P.}~\bibnamefont {Parotto}}, \bibinfo {author} {\bibfnamefont {A.}~\bibnamefont {Pasztor}}, \bibinfo {author} {\bibfnamefont {L.}~\bibnamefont {Pirelli}}, \bibinfo {author} {\bibfnamefont {K.~K.}\ \bibnamefont {Szabo}}, \ and\ \bibinfo {author} {\bibfnamefont {C.~H.}\ \bibnamefont {Wong}},\ }\href {\doibase 10.1103/PhysRevD.110.114507} {\bibfield  {journal} {\bibinfo  {journal} {Phys. Rev. D}\ }\textbf {\bibinfo {volume} {110}},\ \bibinfo {pages} {114507} (\bibinfo {year} {2024}{\natexlab{c}})},\ \Eprint {http://arxiv.org/abs/2410.06216} {arXiv:2410.06216 [hep-lat]} \BibitemShut {NoStop}%
\bibitem [{\citenamefont {Hasenfratz}\ and\ \citenamefont {Toussaint}(1992)}]{Hasenfratz:1991ax}%
  \BibitemOpen
  \bibfield  {author} {\bibinfo {author} {\bibfnamefont {A.}~\bibnamefont {Hasenfratz}}\ and\ \bibinfo {author} {\bibfnamefont {D.}~\bibnamefont {Toussaint}},\ }\href {\doibase 10.1016/0550-3213(92)90247-9} {\bibfield  {journal} {\bibinfo  {journal} {Nucl. Phys.}\ }\textbf {\bibinfo {volume} {B371}},\ \bibinfo {pages} {539} (\bibinfo {year} {1992})}\BibitemShut {NoStop}%
\bibitem [{\citenamefont {Allton}\ \emph {et~al.}(2005)\citenamefont {Allton}, \citenamefont {Doring}, \citenamefont {Ejiri}, \citenamefont {Hands}, \citenamefont {Kaczmarek} \emph {et~al.}}]{Allton:2005gk}%
  \BibitemOpen
  \bibfield  {author} {\bibinfo {author} {\bibfnamefont {C.}~\bibnamefont {Allton}}, \bibinfo {author} {\bibfnamefont {M.}~\bibnamefont {Doring}}, \bibinfo {author} {\bibfnamefont {S.}~\bibnamefont {Ejiri}}, \bibinfo {author} {\bibfnamefont {S.}~\bibnamefont {Hands}}, \bibinfo {author} {\bibfnamefont {O.}~\bibnamefont {Kaczmarek}},  \emph {et~al.},\ }\href {\doibase 10.1103/PhysRevD.71.054508} {\bibfield  {journal} {\bibinfo  {journal} {Phys.Rev.}\ }\textbf {\bibinfo {volume} {D71}},\ \bibinfo {pages} {054508} (\bibinfo {year} {2005})},\ \Eprint {http://arxiv.org/abs/hep-lat/0501030} {arXiv:hep-lat/0501030 [hep-lat]} \BibitemShut {NoStop}%
\bibitem [{Note1()}]{Note1}%
  \BibitemOpen
  \bibinfo {note} {We supply these coefficients as ancillary files along with the submission.}\BibitemShut {Stop}%
\bibitem [{\citenamefont {Stephanov}(2011)}]{Stephanov:2011pb}%
  \BibitemOpen
  \bibfield  {author} {\bibinfo {author} {\bibfnamefont {M.}~\bibnamefont {Stephanov}},\ }\href {\doibase 10.1103/PhysRevLett.107.052301} {\bibfield  {journal} {\bibinfo  {journal} {Phys.Rev.Lett.}\ }\textbf {\bibinfo {volume} {107}},\ \bibinfo {pages} {052301} (\bibinfo {year} {2011})},\ \Eprint {http://arxiv.org/abs/1104.1627} {arXiv:1104.1627 [hep-ph]} \BibitemShut {NoStop}%
\bibitem [{\citenamefont {Mroczek}\ \emph {et~al.}(2021)\citenamefont {Mroczek}, \citenamefont {Nava~Acuna}, \citenamefont {Noronha-Hostler}, \citenamefont {Parotto}, \citenamefont {Ratti},\ and\ \citenamefont {Stephanov}}]{Mroczek:2020rpm}%
  \BibitemOpen
  \bibfield  {author} {\bibinfo {author} {\bibfnamefont {D.}~\bibnamefont {Mroczek}}, \bibinfo {author} {\bibfnamefont {A.~R.}\ \bibnamefont {Nava~Acuna}}, \bibinfo {author} {\bibfnamefont {J.}~\bibnamefont {Noronha-Hostler}}, \bibinfo {author} {\bibfnamefont {P.}~\bibnamefont {Parotto}}, \bibinfo {author} {\bibfnamefont {C.}~\bibnamefont {Ratti}}, \ and\ \bibinfo {author} {\bibfnamefont {M.~A.}\ \bibnamefont {Stephanov}},\ }\href {\doibase 10.1103/PhysRevC.103.034901} {\bibfield  {journal} {\bibinfo  {journal} {Phys. Rev. C}\ }\textbf {\bibinfo {volume} {103}},\ \bibinfo {pages} {034901} (\bibinfo {year} {2021})},\ \Eprint {http://arxiv.org/abs/2008.04022} {arXiv:2008.04022 [nucl-th]} \BibitemShut {NoStop}%
\bibitem [{\citenamefont {Fu}\ \emph {et~al.}(2021)\citenamefont {Fu}, \citenamefont {Luo}, \citenamefont {Pawlowski}, \citenamefont {Rennecke}, \citenamefont {Wen},\ and\ \citenamefont {Yin}}]{Fu:2021oaw}%
  \BibitemOpen
  \bibfield  {author} {\bibinfo {author} {\bibfnamefont {W.-j.}\ \bibnamefont {Fu}}, \bibinfo {author} {\bibfnamefont {X.}~\bibnamefont {Luo}}, \bibinfo {author} {\bibfnamefont {J.~M.}\ \bibnamefont {Pawlowski}}, \bibinfo {author} {\bibfnamefont {F.}~\bibnamefont {Rennecke}}, \bibinfo {author} {\bibfnamefont {R.}~\bibnamefont {Wen}}, \ and\ \bibinfo {author} {\bibfnamefont {S.}~\bibnamefont {Yin}},\ }\href {\doibase 10.1103/PhysRevD.104.094047} {\bibfield  {journal} {\bibinfo  {journal} {Phys. Rev. D}\ }\textbf {\bibinfo {volume} {104}},\ \bibinfo {pages} {094047} (\bibinfo {year} {2021})},\ \Eprint {http://arxiv.org/abs/2101.06035} {arXiv:2101.06035 [hep-ph]} \BibitemShut {NoStop}%
\bibitem [{\citenamefont {Andronic}\ \emph {et~al.}(2006)\citenamefont {Andronic}, \citenamefont {Braun-Munzinger},\ and\ \citenamefont {Stachel}}]{Andronic:2005yp}%
  \BibitemOpen
  \bibfield  {author} {\bibinfo {author} {\bibfnamefont {A.}~\bibnamefont {Andronic}}, \bibinfo {author} {\bibfnamefont {P.}~\bibnamefont {Braun-Munzinger}}, \ and\ \bibinfo {author} {\bibfnamefont {J.}~\bibnamefont {Stachel}},\ }\href {\doibase 10.1016/j.nuclphysa.2006.03.012} {\bibfield  {journal} {\bibinfo  {journal} {Nucl.Phys.}\ }\textbf {\bibinfo {volume} {A772}},\ \bibinfo {pages} {167} (\bibinfo {year} {2006})},\ \Eprint {http://arxiv.org/abs/nucl-th/0511071} {arXiv:nucl-th/0511071 [nucl-th]} \BibitemShut {NoStop}%
\bibitem [{\citenamefont {Becattini}\ \emph {et~al.}(2013)\citenamefont {Becattini}, \citenamefont {Bleicher}, \citenamefont {Kollegger}, \citenamefont {Schuster}, \citenamefont {Steinheimer} \emph {et~al.}}]{Becattini:2012xb}%
  \BibitemOpen
  \bibfield  {author} {\bibinfo {author} {\bibfnamefont {F.}~\bibnamefont {Becattini}}, \bibinfo {author} {\bibfnamefont {M.}~\bibnamefont {Bleicher}}, \bibinfo {author} {\bibfnamefont {T.}~\bibnamefont {Kollegger}}, \bibinfo {author} {\bibfnamefont {T.}~\bibnamefont {Schuster}}, \bibinfo {author} {\bibfnamefont {J.}~\bibnamefont {Steinheimer}},  \emph {et~al.},\ }\href {\doibase 10.1103/PhysRevLett.111.082302} {\bibfield  {journal} {\bibinfo  {journal} {Phys.Rev.Lett.}\ }\textbf {\bibinfo {volume} {111}},\ \bibinfo {pages} {082302} (\bibinfo {year} {2013})},\ \Eprint {http://arxiv.org/abs/1212.2431} {arXiv:1212.2431 [nucl-th]} \BibitemShut {NoStop}%
\bibitem [{\citenamefont {Alba}\ \emph {et~al.}(2014)\citenamefont {Alba}, \citenamefont {Alberico}, \citenamefont {Bellwied}, \citenamefont {Bluhm}, \citenamefont {Mantovani~Sarti} \emph {et~al.}}]{Alba:2014eba}%
  \BibitemOpen
  \bibfield  {author} {\bibinfo {author} {\bibfnamefont {P.}~\bibnamefont {Alba}}, \bibinfo {author} {\bibfnamefont {W.}~\bibnamefont {Alberico}}, \bibinfo {author} {\bibfnamefont {R.}~\bibnamefont {Bellwied}}, \bibinfo {author} {\bibfnamefont {M.}~\bibnamefont {Bluhm}}, \bibinfo {author} {\bibfnamefont {V.}~\bibnamefont {Mantovani~Sarti}},  \emph {et~al.},\ }\href {\doibase 10.1016/j.physletb.2014.09.052} {\bibfield  {journal} {\bibinfo  {journal} {Phys.Lett.}\ }\textbf {\bibinfo {volume} {B738}},\ \bibinfo {pages} {305} (\bibinfo {year} {2014})},\ \Eprint {http://arxiv.org/abs/1403.4903} {arXiv:1403.4903 [hep-ph]} \BibitemShut {NoStop}%
\bibitem [{\citenamefont {Vovchenko}\ \emph {et~al.}(2016)\citenamefont {Vovchenko}, \citenamefont {Begun},\ and\ \citenamefont {Gorenstein}}]{Vovchenko:2015idt}%
  \BibitemOpen
  \bibfield  {author} {\bibinfo {author} {\bibfnamefont {V.}~\bibnamefont {Vovchenko}}, \bibinfo {author} {\bibfnamefont {V.~V.}\ \bibnamefont {Begun}}, \ and\ \bibinfo {author} {\bibfnamefont {M.~I.}\ \bibnamefont {Gorenstein}},\ }\href {\doibase 10.1103/PhysRevC.93.064906} {\bibfield  {journal} {\bibinfo  {journal} {Phys. Rev.}\ }\textbf {\bibinfo {volume} {C93}},\ \bibinfo {pages} {064906} (\bibinfo {year} {2016})},\ \Eprint {http://arxiv.org/abs/1512.08025} {arXiv:1512.08025 [nucl-th]} \BibitemShut {NoStop}%
\bibitem [{\citenamefont {Itzykson}\ \emph {et~al.}(1983)\citenamefont {Itzykson}, \citenamefont {Pearson},\ and\ \citenamefont {Zuber}}]{Itzykson:1983gb}%
  \BibitemOpen
  \bibfield  {author} {\bibinfo {author} {\bibfnamefont {C.}~\bibnamefont {Itzykson}}, \bibinfo {author} {\bibfnamefont {R.~B.}\ \bibnamefont {Pearson}}, \ and\ \bibinfo {author} {\bibfnamefont {J.~B.}\ \bibnamefont {Zuber}},\ }\href {\doibase 10.1016/0550-3213(83)90499-6} {\bibfield  {journal} {\bibinfo  {journal} {Nucl. Phys. B}\ }\textbf {\bibinfo {volume} {220}},\ \bibinfo {pages} {415} (\bibinfo {year} {1983})}\BibitemShut {NoStop}%
\bibitem [{\citenamefont {Butera}\ and\ \citenamefont {Pernici}(2012)}]{Butera:2012tq}%
  \BibitemOpen
  \bibfield  {author} {\bibinfo {author} {\bibfnamefont {P.}~\bibnamefont {Butera}}\ and\ \bibinfo {author} {\bibfnamefont {M.}~\bibnamefont {Pernici}},\ }\href {\doibase 10.1103/PhysRevE.86.011104} {\bibfield  {journal} {\bibinfo  {journal} {Phys. Rev. E}\ }\textbf {\bibinfo {volume} {86}},\ \bibinfo {pages} {011104} (\bibinfo {year} {2012})},\ \Eprint {http://arxiv.org/abs/1206.0872} {arXiv:1206.0872 [cond-mat.stat-mech]} \BibitemShut {NoStop}%
\bibitem [{\citenamefont {Gliozzi}(2013)}]{Gliozzi:2013ysa}%
  \BibitemOpen
  \bibfield  {author} {\bibinfo {author} {\bibfnamefont {F.}~\bibnamefont {Gliozzi}},\ }\href {\doibase 10.1103/PhysRevLett.111.161602} {\bibfield  {journal} {\bibinfo  {journal} {Phys. Rev. Lett.}\ }\textbf {\bibinfo {volume} {111}},\ \bibinfo {pages} {161602} (\bibinfo {year} {2013})},\ \Eprint {http://arxiv.org/abs/1307.3111} {arXiv:1307.3111 [hep-th]} \BibitemShut {NoStop}%
\bibitem [{\citenamefont {Vovchenko}\ \emph {et~al.}(2017{\natexlab{a}})\citenamefont {Vovchenko}, \citenamefont {Pasztor}, \citenamefont {Fodor}, \citenamefont {Katz},\ and\ \citenamefont {Stoecker}}]{Vovchenko:2017xad}%
  \BibitemOpen
  \bibfield  {author} {\bibinfo {author} {\bibfnamefont {V.}~\bibnamefont {Vovchenko}}, \bibinfo {author} {\bibfnamefont {A.}~\bibnamefont {Pasztor}}, \bibinfo {author} {\bibfnamefont {Z.}~\bibnamefont {Fodor}}, \bibinfo {author} {\bibfnamefont {S.~D.}\ \bibnamefont {Katz}}, \ and\ \bibinfo {author} {\bibfnamefont {H.}~\bibnamefont {Stoecker}},\ }\href {\doibase 10.1016/j.physletb.2017.10.042} {\bibfield  {journal} {\bibinfo  {journal} {Phys. Lett.}\ }\textbf {\bibinfo {volume} {B775}},\ \bibinfo {pages} {71} (\bibinfo {year} {2017}{\natexlab{a}})},\ \Eprint {http://arxiv.org/abs/1708.02852} {arXiv:1708.02852 [hep-ph]} \BibitemShut {NoStop}%
\bibitem [{\citenamefont {Huovinen}\ and\ \citenamefont {Petreczky}(2018)}]{Huovinen:2017ogf}%
  \BibitemOpen
  \bibfield  {author} {\bibinfo {author} {\bibfnamefont {P.}~\bibnamefont {Huovinen}}\ and\ \bibinfo {author} {\bibfnamefont {P.}~\bibnamefont {Petreczky}},\ }\href {\doibase 10.1016/j.physletb.2017.12.001} {\bibfield  {journal} {\bibinfo  {journal} {Phys. Lett.}\ }\textbf {\bibinfo {volume} {B777}},\ \bibinfo {pages} {125} (\bibinfo {year} {2018})},\ \Eprint {http://arxiv.org/abs/1708.00879} {arXiv:1708.00879 [hep-ph]} \BibitemShut {NoStop}%
\bibitem [{\citenamefont {Bellwied}\ \emph {et~al.}(2021)\citenamefont {Bellwied}, \citenamefont {Borsanyi}, \citenamefont {Fodor}, \citenamefont {Guenther}, \citenamefont {Katz}, \citenamefont {Parotto}, \citenamefont {Pasztor}, \citenamefont {Pesznyak}, \citenamefont {Ratti},\ and\ \citenamefont {Szabo}}]{Bellwied:2021nrt}%
  \BibitemOpen
  \bibfield  {author} {\bibinfo {author} {\bibfnamefont {R.}~\bibnamefont {Bellwied}}, \bibinfo {author} {\bibfnamefont {S.}~\bibnamefont {Borsanyi}}, \bibinfo {author} {\bibfnamefont {Z.}~\bibnamefont {Fodor}}, \bibinfo {author} {\bibfnamefont {J.~N.}\ \bibnamefont {Guenther}}, \bibinfo {author} {\bibfnamefont {S.~D.}\ \bibnamefont {Katz}}, \bibinfo {author} {\bibfnamefont {P.}~\bibnamefont {Parotto}}, \bibinfo {author} {\bibfnamefont {A.}~\bibnamefont {Pasztor}}, \bibinfo {author} {\bibfnamefont {D.}~\bibnamefont {Pesznyak}}, \bibinfo {author} {\bibfnamefont {C.}~\bibnamefont {Ratti}}, \ and\ \bibinfo {author} {\bibfnamefont {K.~K.}\ \bibnamefont {Szabo}},\ }\href {\doibase 10.1103/PhysRevD.104.094508} {\bibfield  {journal} {\bibinfo  {journal} {Phys. Rev. D}\ }\textbf {\bibinfo {volume} {104}},\ \bibinfo {pages} {094508} (\bibinfo {year} {2021})},\ \Eprint {http://arxiv.org/abs/2102.06625} {arXiv:2102.06625 [hep-lat]} \BibitemShut {NoStop}%
\bibitem [{\citenamefont {Roberge}\ and\ \citenamefont {Weiss}(1986)}]{Roberge:1986mm}%
  \BibitemOpen
  \bibfield  {author} {\bibinfo {author} {\bibfnamefont {A.}~\bibnamefont {Roberge}}\ and\ \bibinfo {author} {\bibfnamefont {N.}~\bibnamefont {Weiss}},\ }\href {\doibase 10.1016/0550-3213(86)90582-1} {\bibfield  {journal} {\bibinfo  {journal} {Nucl.Phys.}\ }\textbf {\bibinfo {volume} {B275}},\ \bibinfo {pages} {734} (\bibinfo {year} {1986})}\BibitemShut {NoStop}%
\bibitem [{\citenamefont {Elliott}\ \emph {et~al.}(2013)\citenamefont {Elliott}, \citenamefont {Lake}, \citenamefont {Moretto},\ and\ \citenamefont {Phair}}]{Elliott:2013pna}%
  \BibitemOpen
  \bibfield  {author} {\bibinfo {author} {\bibfnamefont {J.~B.}\ \bibnamefont {Elliott}}, \bibinfo {author} {\bibfnamefont {P.~T.}\ \bibnamefont {Lake}}, \bibinfo {author} {\bibfnamefont {L.~G.}\ \bibnamefont {Moretto}}, \ and\ \bibinfo {author} {\bibfnamefont {L.}~\bibnamefont {Phair}},\ }\href {\doibase 10.1103/PhysRevC.87.054622} {\bibfield  {journal} {\bibinfo  {journal} {Phys. Rev. C}\ }\textbf {\bibinfo {volume} {87}},\ \bibinfo {pages} {054622} (\bibinfo {year} {2013})}\BibitemShut {NoStop}%
\bibitem [{\citenamefont {Vovchenko}\ \emph {et~al.}(2017{\natexlab{b}})\citenamefont {Vovchenko}, \citenamefont {Gorenstein},\ and\ \citenamefont {Stoecker}}]{Vovchenko:2016rkn}%
  \BibitemOpen
  \bibfield  {author} {\bibinfo {author} {\bibfnamefont {V.}~\bibnamefont {Vovchenko}}, \bibinfo {author} {\bibfnamefont {M.~I.}\ \bibnamefont {Gorenstein}}, \ and\ \bibinfo {author} {\bibfnamefont {H.}~\bibnamefont {Stoecker}},\ }\href {\doibase 10.1103/PhysRevLett.118.182301} {\bibfield  {journal} {\bibinfo  {journal} {Phys. Rev. Lett.}\ }\textbf {\bibinfo {volume} {118}},\ \bibinfo {pages} {182301} (\bibinfo {year} {2017}{\natexlab{b}})},\ \Eprint {http://arxiv.org/abs/1609.03975} {arXiv:1609.03975 [hep-ph]} \BibitemShut {NoStop}%
\end{thebibliography}
%

\end{document}